\documentclass[11pt,a4paper]{article}

\usepackage[T1]{fontenc}
\usepackage{lmodern}
\usepackage{amsmath,amssymb,amsthm}
\usepackage[margin=1.1in]{geometry}
\usepackage[authoryear]{natbib}
\setcitestyle{authoryear,open={(},close={)}}
\usepackage{comment}

\theoremstyle{plain}
\newtheorem{theorem}{Theorem}[section]
\newtheorem{proposition}[theorem]{Proposition}
\newtheorem{lemma}[theorem]{Lemma}
\newtheorem{corollary}[theorem]{Corollary}
\theoremstyle{definition}
\newtheorem{example}[theorem]{Example}
\newtheorem{remark}[theorem]{Remark}
\usepackage{microtype}
\usepackage{xurl}
\usepackage{graphicx}
\usepackage{authblk}

\usepackage[hidelinks,hypertexnames=false,hyperfootnotes=false]{hyperref}
\hypersetup{pdftitle={Path-Independent Deleveraging Rules},pdfauthor={Christoph Schlegel}}

\newcommand{\R}{\mathbb R}
\newcommand{\Rpos}{\R_{>0}}
\newcommand{\id}{\operatorname{id}}

\title{Path-Independent Deleveraging Rules}
\author{Christoph Schlegel and Akaki Mamageishvili}
\affil{Flashbots and Offchain}
\date{}

\begin{document}
\maketitle

\begin{abstract}
We characterise path-independent auto-deleveraging (ADL) rules as an extended log-exponential family. Immunity to account splitting, merging and reopening further restrict the class. Homogeneity (essentially) restricts the class to a one-parameter family ordered by both fairness and platform risk. Its split-proof members continuously interpolate between proportional division and water-filling. Among these rules, decreasing the parameter reduces risk at the cost of increasing inequality in how accounts are treated. This suggests selecting the parameter through an ex-ante calibration to a risk bound on historical data to choose a maximally fair rule subject to that bound. A counterfactual replay of Hyperliquid's ADL episode on 10 October 2025 finds that a range of intermediate parameters significantly improves on the observed allocation in both burden dispersion and subsequent risk statistics.
\end{abstract}

\section{Introduction}

Perpetual futures are widely used to take and hedge directional exposure to cryptoassets. They provide capital-efficient long and short exposure without expiry, using leverage, and perpetual futures exchanges contribute to price discovery of their underlying assets~\citep{alexanderHeck2020,chitraOnlineADL2026}. Leverage also creates a solvency problem. If a losing account is liquidated only after the market has moved through its bankruptcy price, its collateral no longer covers the loss. Insurance funds absorb ordinary shortfalls; when they are insufficient, auto-deleveraging (ADL) forcibly reduces profitable positions on the opposite side. 
A roughly 30\% five-minute Bitcoin price decline in March 2017 exhausted BitMEX's insurance fund and triggered ADL of profitable shorts; further ADL occurred in BitMEX's XBT and ETH futures in April 2019; and
ADL was activated during Hyperliquid's 10 October 2025 stress episode~\citep{bitmexInsurance2019,bitmexADL2019,chitraOnlineADL2026}.

ADL is a more prominent part of market design in crypto than the analogous partial tear-up of traditional futures positions. Cryptoassets are more volatile than traditional financial assets and crypto derivatives commonly permit high leverage~\citep{bisCryptoRisk2023,chitraOnlineADL2026}. Crypto venues also rely more directly on posted collateral, automated liquidation and insurance funds than traditional central counterparties, which interpose a clearing layer backed by margin, default funds and other recovery resources~\citep{cpmiIOSCO2014}; on-chain venues in particular are designed to operate without a traditional clearing-member intermediary~\citep{harveyHasbrouckSaleh2026}. Reduced reliance on intermediation is part of the appeal of this architecture, but makes the allocation of residual losses more important.

How ADL is done matters because it acts on traders who are solvent and usually profitable. A forced reduction can remove a hedge and change a trader's exposure precisely during stressed markets. The rule therefore determines who bears the burden, how predictable that exposure is ex ante, whether treatment can be manipulated by restructuring accounts, and how much risk remains after deleveraging. These objectives do not necessarily coincide. Proportional reductions are simple and predictable but do not prioritise the riskiest positions; risk-based rules can reduce platform risk but may concentrate the burden on a small set of traders~\citep{campbell,chitraADL2025}. Existing venues indeed use different policies, including rankings based on leverage, profitability, or both, as well as random selection~\citep{hyperliquid2026,lighter2026,aster2026,extended2026,edgex2026,dydx2026}. There is therefore no settled convention for how the ADL burden should be allocated. 

We take an axiomatic approach to the design of ADL rules, characterising the class of rules that satisfy a set of basic requirements, and trade-off risk and fairness to select a rule within this class. The starting point of the characterization is path independence. Call the aggregate open interest that remains after ADL the target. If the target is revised downward, applying the rule to the surviving positions should agree with applying it directly to the final target; if it is revised upward, restoring positions in stages should agree with allocating the larger target directly. Together with continuity, equal treatment of equals and consistency, path independence characterises an extended log-exponential family.

We then ask whether traders can improve their treatment by reorganising the same position and equity across accounts: split-proofness and merge-proofness further restrict the characterised family through superadditivity and subadditivity of its residual maps, respectively, and proportional division is the only member satisfying both.\footnote{\citet{campbell} use the term ``Sybil resistant'' for split-proofness.} In practice, some ADL policies use unrealised P\&L and hence entry prices as an input. Our rule design deliberately ignores entry prices. Section~\ref{sec:reopening} gives a manipulation-based reason for this restriction: closing and immediately reopening a position changes its entry price without changing its current position or equity, and we show that preventing this operation from improving later ADL treatment is equivalent to entry-price independence.

Homogeneity requires allocations to scale when positions, equities and the target are scaled by a common factor. It selects a one-parameter family of finite rules and allows additional boundary rules. Section~\ref{sec:fairness-risk} explains why we focus on the finite family and its limits.

The rules between proportional division and water-filling\footnote{Water-filling reduces positions whose leverage exceeds a common threshold to that threshold and leaves the others unchanged, with the threshold chosen to meet the aggregate target.} in the one-parameter family offer a trade-off between fairness and risk. Moving towards water-filling lowers risk but makes percentage reductions less equal. 
We also characterise the two endpoints of this trade-off more generally. Proportional division uniquely gives the fairest allocation in the position-weighted Lorenz order on percentage reductions, while water-filling minimises platform risk state by state. This gives a simple one-parameter menu for rule design when homogeneity is acceptable and a platform can use historical data to select a parameter from the menu. A natural calibration is to minimize dispersion subject to a specified risk bound being met on that historical data.
When heterogeneity is desirable, the design problem is less tidy, but our general results continue to apply.

We illustrate the trade-off with a counterfactual replay of Hyperliquid's ADL episode on 10 October 2025 in Section~\ref{sec:empirical}. Across 18 markets, every evaluated parameter 
below a certain threshold, including water-filling, improves on the observed allocation in the reported fairness and risk statistics. Intermediate parameters distribute percentage reductions more equally than water-filling while retaining lower risk than the observed allocation. 

\paragraph{Related literature.}
Formal work on ADL studies loss socialisation, solvency and fairness~\citep{chitraADL2025}, sequential implementation \citep{chitraOnlineADL2026}, and risk minimisation~\citep{campbell}. In the single-asset isolated-margin model, water-filling minimises realised loss at every terminal price over the full feasible set and hence every monotone risk measure \citep[Theorem~2.4(d) and Proposition~2.18]{campbell}; it is weighted {\it constrained equal awards} (CEA) in residual-position space~\citep[Appendix~A.4]{campbell}.\footnote{The pointwise risk optimality of water-filling and its weighted-CEA interpretation are due to \citet{campbell}. Theorem~\ref{thm:risk-optimal-composition} below is instead a rule-level uniqueness result: for a fixed spectral risk criterion, universal optimality within our characterised class selects water-filling.}

Central counter-parties (CCP) recovery provides the closest traditional-finance analogue. After prefunded resources and auctions are exhausted, recovery may use assessments, variation-margin gains haircutting and partial tear-up~\citep{cpmiIOSCO2014,berndsenRecovery2023}. Partial tear-up is especially close to ADL because it involuntarily terminates positions to restore a matched book, whereas variation-margin haircutting allocates losses without reducing positions; the 1989 New Zealand futures clearing crisis is a historical example~\citep{coxMurphyBudding2016}.

When equity is suppressed, ADL is a conflicting-claims problem: positions are claims, surviving open interest is the estate, and residual positions are awards. Proportional division, CEA and {\it constrained equal losses} (CEL) are classical claims rules. Consistency is the standard reduced-problem requirement, while composition down and composition up are the claims-theoretic counterparts of target path independence. The corresponding literature provides both the axiomatic language and the parametric machinery used below~\citep{oneillTalmud1982,aumannMaschler1985,young1987,kaminski2006,moulinCostSharing2002,thomsonClaims2003,thomsonClaims2015,thomsonDivide2019}.

\citet{moulinPriority2000} studies consistency together with upper and lower composition and characterises asymmetric rationing rules under scale invariance. \citet{chambersAsymmetric2006} removes homogeneity, characterises the two composition axioms through difference rules, and constructs consistent logarithmic-proportional rules. These are direct antecedents of our characterisation; here equity is an additional account characteristic on which treatment may depend continuously.

\citet{schlegel} characterises bi-compositional division rules for one-dimensional claimant types, where the claim (position) is the only characteristic and equal claims receive equal awards. Our characterisation extends this result to accounts described by both position and equity. We require equal treatment only when both coordinates agree, allowing accounts with equal positions to receive different treatment according to their equities. In the finite construction, each account may therefore use a different (extended) log-exponential curve where the curve parameter is a function of equity, and the residual positions are determined at a common clearing level.

\section{Model and characterisation}\label{sec:model}

\subsection{Accounts, rules and path independence}

We consider positions on one side, long or short, of a single-asset market. An account $i$ has \textbf{type} $(q_i,e_i)\in\R_+\times\Rpos$, where $q_i$ is its \textbf{position size} in units of the asset and $e_i$ its \textbf{equity} in dollars, valued at the \textbf{mark price} $p_\tau$ at which ADL executes, as in \citet{campbell}. Its \textbf{leverage} is $p_\tau q_i/e_i$, notional over equity. Section~\ref{sec:manipulation} justifies omitting entry prices by requiring that closing and reopening a position cannot improve its ADL treatment. For a finite \textbf{population} $N$, an ADL \textbf{problem} consists of a \textbf{type-profile} $(q,e)=((q_i,e_i))_{i\in N}$ and a \textbf{target} surviving open interest $\widehat E\in[0,\sum_iq_i]$. A \textbf{rule} $r$ is defined for every finite population and assigns to each problem $(q,e,\widehat E)$ a vector of \textbf{residual positions} $r(q,e,\widehat E)=(r_i)_{i\in N}$ satisfying
\[
 0\le r_i\le q_i,\qquad \sum_i r_i=\widehat E.
\]
The forced \textbf{reduction} of account $i$ is $q_i-r_i$. For shorts, this is a forced buyback.\footnote{Reductions and restorations execute at the prevailing mark price
$p_\tau$. We value equity at this price, ignore transaction costs and
funding, and retain realised P\&L.
On Hyperliquid, ADL executes at the previous mark price, so $p_\tau$ is
the mark at which the exchange measures equity.}

We impose the following regularity and invariance requirements on rules.

\medskip\noindent\textbf{Equal treatment of equal accounts.}
Accounts with the same type receive the same residual position.

\medskip\noindent\textbf{Continuity.}
The rule is continuous in positions, equities and the target for each population.

\medskip\noindent\textbf{Consistency.}
Fix a problem $(q,e,\widehat E)$ and write $r^*=r(q,e,\widehat E)$. For every nonempty subpopulation $S\subseteq N$, let $q_S=(q_i)_{i\in S}$ and $e_S=(e_i)_{i\in S}$. Consistency requires

\begin{equation}\label{eq:consistency}
 r\left(q_S,e_S,\sum_{i\in S}r_i^*\right)=r_S^*
 \qquad\text{for every nonempty }S\subseteq N.
\end{equation}

Thus, if the accounts outside $S$ leave with exactly the residual positions assigned to them by the rule, reassessing the problem for the accounts that remain does not change their allocation.

Consistency is a standard reduced-problem axiom in claims and resource allocation.~\citet{young1987} describes it as a basic principle, and it is a central variable-population requirement in Thomson's surveys and monograph \citep{thomsonClaims2003,thomsonConsistency2012,thomsonDivide2019}. It can be interpreted as path independence in population space. It rules out procedures that depend arbitrarily on population size, for example using water-filling for an odd number of accounts and proportional division for an even number.

Consistency compares allocations before and after reducing the population; path independence compares allocations obtained by adjusting the target in stages or directly. These must agree whether the target falls or rises, keeping equities fixed.

\medskip\noindent\textbf{Downward path independence (DPI).}
For $0\le\widehat E'\le\widehat E\le\sum_iq_i$,
\[
 r\bigl(r(q,e,\widehat E),e,\widehat E'\bigr)
 =r(q,e,\widehat E').
\]
If the target is tightened from $\widehat E$ to $\widehat E'$, applying the rule again to the surviving positions gives the same final allocation as applying the tighter target directly.

\medskip\noindent\textbf{Upward path independence (UPI).}
For $0\le\widehat E\le\widehat E'\le\sum_iq_i$,
\[
 r(q,e,\widehat E)+
 r\bigl(q-r(q,e,\widehat E),e,\widehat E'-\widehat E\bigr)
 =r(q,e,\widehat E').
\]
If the target increases from $\widehat E$ to $\widehat E'$, restoring $\widehat E'-\widehat E$ of open interest from the positions reduced at $\widehat E$ should give the allocation at $\widehat E'$.

A rule is \textbf{path independent} if it satisfies both DPI and UPI.

\subsection{The extended log-exponential family}

The rules in our characterisation assign each account a \textbf{residual-position schedule} $F_\lambda(q_i,e_i)$, depending on its type and a common \textbf{clearing level} $\lambda\geq0$. For a given problem, this level is chosen to satisfy the target:
\[
 r_i=F_\lambda(q_i,e_i),\qquad \sum_iF_\lambda(q_i,e_i)=\widehat E.
\]
The rule fixes the schedules in advance; only the clearing level adjusts to the problem.\footnote{\label{fn:lambda-param}The common clearing level is an auxiliary parameter. A common strictly increasing relabelling of all residual schedules leaves the rule unchanged. The CEL boundary is written in terms of a common loss level, which decreases as the target increases; reversing its orientation gives an increasing residual parametrisation.}

For $b\in\R$, define
\[
 \varphi_b(x):=
 \begin{cases}
  (\exp(bx)-1)/b,&b\ne0,\\
  x,&b=0.
 \end{cases}
\]
A continuous \textbf{parameter profile} $b:\Rpos\to\R$ specifies a \textbf{finite log-exponential rule} through
\[
F_\lambda(q_i,e_i)=\varphi_{b(e_i)}^{-1}(\lambda\varphi_{b(e_i)}(q_i)),\qquad 0\le\lambda\le1.
\]
At a fixed equity and an interior level, $b(x)<0$ gives smaller positions a larger retained percentage; $b(x)>0$ gives larger positions a larger retained percentage. For $b\equiv0$, clearing gives the \textbf{proportional rule}
\[
 r_i^{\rm prop}(q,e,\widehat E)=\frac{\widehat E}{\sum_jq_j}\,q_i
\]
when total position is positive. Under proportional division, every account with a positive position retains the same percentage of that position.

We also define the \textbf{weighted constrained equal awards} (CEA) and \textbf{weighted constrained equal losses} (CEL) rules. For $w\geq0 $ and $\lambda\geq0$ define
\[
 \mathrm{CEA}_w(x,\lambda)=\min\{x,w\lambda\},
 \qquad
 \mathrm{CEL}_w(x,\lambda)=\max\{0,x-w\lambda\}.
\]
A positive continuous \textbf{weight profile} $w:\Rpos\to\Rpos$ specifies a \textbf{boundary rule} through $$F_\lambda(q_i,e_i)=\text{CEA}_{w(e_i)}(q_i,\lambda)\quad \text{resp.}\quad F_\lambda(q_i,e_i)=\text{CEL}_{w(e_i)}(q_i,\lambda).$$ 

\begin{figure}[t]
\centering
\includegraphics[width=\textwidth]{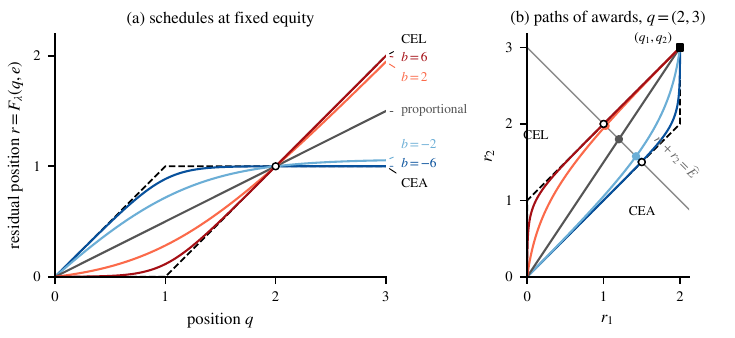}
\caption{The extended log-exponential family.
(a) Residual schedules $r=F_\lambda(q,x)$ at a fixed equity $x$, with $b=b(x)$,
for $b\in\{-6,-2,0,2,6\}$ and the boundary rules $\mathrm{CEA}$ and $\mathrm{CEL}$
(dashed). Clearing levels are normalised so that every schedule passes through
$(q,r)=(2,1)$. For $b<0$, residual positions approach a cap as $q$ grows;
for $b>0$, reductions approach a constant; $b=0$ is proportional.
(b) Paths of awards for two accounts of equal equity with $q=(2,3)$, traced as
the target $\widehat E$ runs from $0$ to $q_1+q_2$ (same colours as in (a)).
Each path runs from the origin to $(q_1,q_2)$; the target selects its
intersection with the budget line $r_1+r_2=\widehat E$, shown for
$\widehat E=3$. Finite-parameter paths approach the CEA and CEL paths
as $|b|$ grows, but do not coincide with them away from the endpoints.}
\label{fig:family}
\end{figure}

\begin{remark}[Boundary rules as limits]
\label{rem:type-endpoint-limits}
Fix a positive continuous weight profile $w$ and consider
$b_\theta(x)=\theta/w(x)$, $\theta\in\R$. Write
\[
 F^\theta_t(q,x):=
 \varphi_{b_\theta(x)}^{-1}\bigl(t\varphi_{b_\theta(x)}(q)\bigr),
 \qquad 0\le t\le1,
\]
for the corresponding residual schedule. Here $\theta$ indexes the
parameter profile, $x>0$ is equity, and $t$ is the clearing level.
After rescaling this level, the schedules converge to the boundary
schedules: for $q,\lambda\ge0$,
\[
 F^\theta_{1-\exp(\theta\lambda)}(q,x)
 \xrightarrow{\theta\to-\infty}\min\{q,w(x)\lambda\},
 \qquad
 F^\theta_{\exp(-\theta\lambda)}(q,x)
 \xrightarrow{\theta\to+\infty}\max\{0,q-w(x)\lambda\}.
\]
The differences between the finite log-exponential rule and the limit rules are bounded by $w(x)\log 2/|\theta|$, uniformly in
$q,\lambda\ge0$. Convergence is therefore uniform on compact equity
sets as well. For each fixed problem, the budget-clearing allocations
converge to those of $\mathrm{CEA}_w$ and $\mathrm{CEL}_w$, respectively.
The relative rates of divergence retain the weight profile $w$.
\end{remark}
Weighted CEA equalises residual positions per unit of weight, subject to the original position caps. With $w(x)=x$, this is \textbf{water-filling}~\citep{campbell}: accounts whose initial leverage exceeds $p_\tau\lambda$ are reduced to leverage $p_\tau\lambda$, while the others retain their positions. Weighted CEL equalises reductions per unit of weight, subject to nonnegative residuals. These boundary rules and the finite rules form the \textbf{extended log-exponential family}, illustrated in Figure~\ref{fig:family}.

The following theorem extends the scalar characterisation in \citet[Theorem~2.2]{schlegel} to equity-dependent rules. The special case with equal equity for all accounts gives the scalar setting. To go from equal to different equity levels, one needs to align the clearing levels: path independence forces a common normalisation and excludes mixtures of finite and boundary rules, while continuity gives a continuous profile $b$ or $w$. This is proved in Appendix~\ref{app:characterisation}. 

\begin{theorem}[Characterisation]
\label{thm:type-comppair}
A rule satisfies equal treatment of equal accounts, continuity, consistency and path independence
if and only if it belongs to the extended log-exponential family.
\end{theorem}

The axioms are
logically independent as we show in
Appendix~\ref{app:independence}.

\subsection{Homogeneity}
The characterisation allows any continuous profile $b$, or any positive continuous profile $w$, so it gives little guidance on how treatment should vary with equity. Homogeneity restricts this dependence on equity by requiring allocations to scale with positions, equities and the target. It leaves one scalar parameter in each of the three families.

\medskip\noindent\textbf{Homogeneity.}
For every problem $(q,e,\widehat E)$ and every common scale factor $t>0$, homogeneity requires
\[
 r(tq,te,t\widehat E)=t\,r(q,e,\widehat E).
\]
The same allocation principle therefore applies at every absolute scale, excluding fixed absolute thresholds. Such thresholds may be relevant when, for example, market depth makes the absolute size of a position matter.

\begin{proposition}[Homogeneous rules]\label{prop:homogeneous}
Within the extended log-exponential family, a finite rule is homogeneous if and only if there exists a constant $\theta\in\R$ such that
\[
 b(x)=\frac{\theta p_\tau}{x}\qquad\text{for every }x>0.
\]
A boundary rule is homogeneous if and only if it admits a weight profile of the form
\[
 w(x)=x^\gamma\qquad\text{for every }x>0,
\]
for some constant $\gamma\in\R$.
\end{proposition}

The proof is in Appendix~\ref{app:splitting}. As noted in footnote~\ref{fn:lambda-param}, the clearing-level label is only a parametrisation. For boundary rules, replacing $w$ by $Cw$ and the clearing level by $\lambda/C$ leaves every product $w(e_i)\lambda$ unchanged. Hence $Cx^\gamma$ and $x^\gamma$ induce the same rule. The constant of a homogeneous finite rule is written as $\theta p_\tau$ so that $b(e_i)q_i=\theta\,p_\tau q_i/e_i$: the parameter $\theta$ is a pure number that multiplies leverage. We define the one-parameter \textbf{$\theta$ family} with $\theta\in[-\infty,+\infty]$ as follows:
for finite $\theta$, its profile is $b(x)=\theta p_\tau/x$,
$r^{-\infty}$ is water-filling i.e. CEA with $w(x)=x$, and $r^{+\infty}$ is CEL with $w(x)=x$.
These boundary rules are the limits in Remark~\ref{rem:type-endpoint-limits} for the weight profile $x/p_\tau$ (or equivalently $x$). 
Section~\ref{sec:fairness-risk} will motivate focusing on the $r^{\theta}$ sub-family by comparing fairness and platform risk across deleveraging problems.
Remark~\ref{rem:homogeneous-boundary-risk} discusses the remaining homogeneous boundary rules and their risk comparisons.
\begin{figure}[t]
\centering
\includegraphics[width=\textwidth]{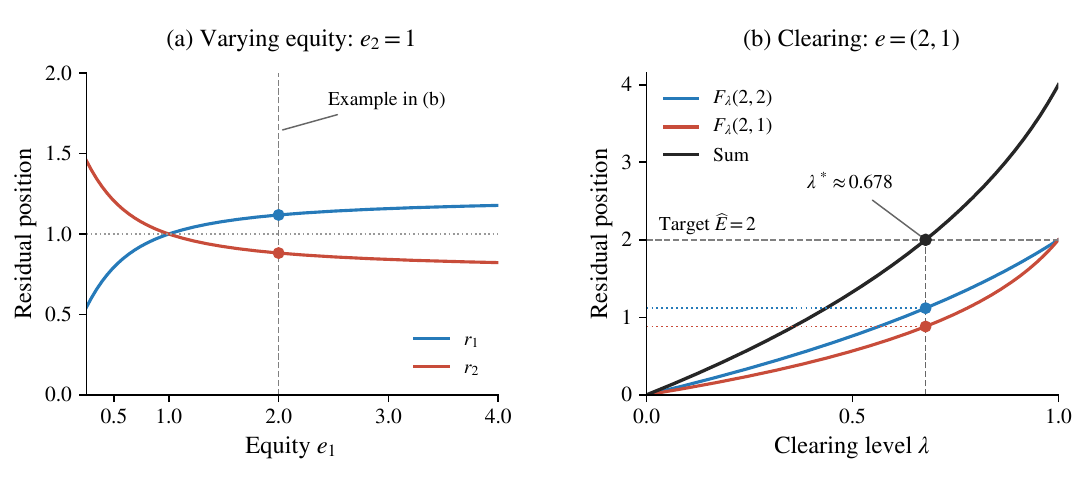}
\caption{Equity dependence and clearing under $r^\theta$ with $\theta=-1$,
$p_\tau=1$, $q=(2,2)$ and $\widehat E=2$.
(a) Residual positions as $e_1$ varies, holding $e_2=1$ fixed and clearing
separately for each profile. Equal equities give $r_1=r_2=1$.
(b) Clearing for $e=(2,1)$, marked by the vertical dashed line in (a).
The individual schedules are
$F_\lambda(2,e_i)=-e_i\log(1-\lambda(1-\exp(-2/e_i)))$.
Their sum meets the target at $\lambda^*\approx0.678$; evaluating both
schedules at this level gives $r_1\approx1.118$ and $r_2\approx0.882$.
The dotted lines mark these residual positions.}
\label{fig:equity}
\end{figure}

\paragraph{The $\theta$-family in leverage space.}\label{rem:leverage-space}
Under each rule $r^\theta$, accounts with the same initial leverage face the same percentage reduction of their positions. The rules $r^\theta$ therefore satisfy a stronger version of equal treatment of equal accounts by extending it to accounts of different sizes. Appendix~\ref{sec:risk-contributions} shows that accounts with equal initial leverage contribute the same risk per unit of equity. They therefore also have equal risk per unit of equity after ADL, i.e.~accounts contributing the same risk per unit of equity are treated equally.
Write $\ell_i=p_\tau q_i/e_i$ and $\widehat\ell_i=p_\tau r_i/e_i$ for leverage, notional over equity, before and after deleveraging~\citep{campbell}. For $b(x)=\theta p_\tau/x$,
\[
 \varphi_{\theta p_\tau/x}(q)=\frac{x}{p_\tau}\,\varphi_\theta\!\left(\frac{p_\tau q}{x}\right),
\]
so $\widehat\ell_i=\varphi_\theta^{-1}(\lambda\varphi_\theta(\ell_i))$. At the endpoints, with the clearing level measured in units of leverage,
\[
 \widehat\ell_i=\min\{\ell_i,\lambda\}\quad\text{for CEA},
 \qquad
 \widehat\ell_i=\max\{0,\ell_i-\lambda\}\quad\text{for CEL},
\]
with $\sum_i e_i\widehat\ell_i=p_\tau\widehat E$, the target notional. Thus $r^\theta$ acts directly on leverage, with equity weights in the budget equation.

Figure~\ref{fig:equity} illustrates the dependence on equity for $\theta=-1$.
With equal initial positions and a fixed target, increasing the
first account's equity increases its residual position. For each equity
profile, the common clearing level is chosen anew to meet the target.

\section{Manipulation}\label{sec:manipulation}

Traders can change how an exposure is recorded without changing the exposure itself. Closing and reopening resets the entry price; splitting and merging redistribute the same aggregate position and equity across accounts. The first operation bears on which account characteristics a rule may use. The second restricts how a rule may depend on positions and equities.

\subsection{Closing and reopening positions}\label{sec:reopening}

So far, an account's type has consisted of its current position and equity. Some deployed ADL policies also use unrealised P\&L, which depends on the entry price. Closing and immediately reopening a position resets that price while, ignoring fees, preserving both position and equity. We ask whether an ADL rule can use entry prices without allowing this operation to improve later treatment.

For short positions, let $s_i>0$ be the \textbf{entry price}. At current price $p$, account $i$ has position $q_i\ge0$, equity $e_i>0$, and cash \textbf{margin}
$m_i=e_i-q_i(s_i-p)>0$. Write
$r(q,e,s;p,\widehat E)$ for the residual positions at target
$\widehat E$. The rule uses these current records and no separate
record of past trades. Consistency and equal treatment of equal accounts apply to the
enlarged type $(q,e,s)$; path independence holds $p,e,s$ fixed.

Suppose account $i$ closes and immediately reopens the same position
at price $p$, replacing $s_i$ by $p$. If the price later moves to
$p'$, both the holding and reopening strategies have positions $q$
and equities
\[
 e'_j=e_j-q_j(p'-p),\qquad j\in N.
\]
Write $s^{i\to p}$ for the entry-price profile after reopening.
We consider prices for which both strategies remain admissible.

\medskip\noindent\textbf{Reopening-proofness.}
For every account $i$, later price $p'$, and target
$\widehat E$,
\begin{equation}\label{eq:reopening-axiom}
 r_i(q,e',s^{i\to p};p',\widehat E)
 \le r_i(q,e',s;p',\widehat E).
\end{equation}
The comparison is required at every admissible later price, whether above or below the reopening price $p$. Under consistency, DPI and equal treatment, it forces the rule to ignore entry prices altogether.

\begin{proposition}[Entry-price independence]\label{thm:entry-irrelevance}
A consistent rule satisfying DPI and equal treatment of equal accounts is reopening-proof if and only if its allocations are independent of entry prices.
\end{proposition}

The proof is in Appendix~\ref{app:reopening}.

\subsection{Splitting and merging accounts}\label{sec:splitting}
A trader may hold the same aggregate position and equity in one account or distribute them across several accounts. We now ask when this choice can improve the trader's ADL allocation. Manipulation by merging or splitting claims is studied by \citet{deFrutos1999}, including for parametric rules. \citet*{juMiyagawaSakai2007} study non-manipulable division with possibly multidimensional characteristics. Our restrictions apply to the path-independent family when both positions and equities are divided or combined.

A rule is \textbf{split-proof} if replacing one account by any finite collection of accounts whose positions and equities sum to those of the original account can never increase the sum of their residual positions. It is \textbf{merge-proof} if consolidating any finite collection of accounts into their sum type can never increase the residual position of the consolidated account.

In the following, for CEL, we reparametrise the level by $-\lambda$, so that all schedules are nondecreasing in the level.

\begin{lemma}[Manipulation and type additivity]\label{lem:typeadd}
Within the extended log-exponential family, a rule is split-proof if and only if every residual map $F_\lambda$ is superadditive in the type,
\[
 F_\lambda(q_1+q_2,e_1+e_2)
 \ge F_\lambda(q_1,e_1)+F_\lambda(q_2,e_2),
\tag{S}
\]
and it is merge-proof if and only if every $F_\lambda$ is subadditive in the type,
\[
 F_\lambda(q_1+q_2,e_1+e_2)
 \le F_\lambda(q_1,e_1)+F_\lambda(q_2,e_2).
\tag{M}
\]
\end{lemma}

\begin{proof}
Suppose (S) holds, and let $\lambda$ clear a problem before a merge.
At that level, merging weakly increases total residual positions. The
merged problem has a clearing level no larger than $\lambda$.
Unaffected accounts receive weakly less, so budget balance gives the
merged account at least the original pair's total. Splitting cannot help.
Reversing the inequalities proves sufficiency of (M).

For necessity, suppose (S) fails for types $z_i=(q_i,e_i)$ at $\lambda$, and let
$S=F_\lambda(z_1)+F_\lambda(z_2)$. By continuity, there are $\mu>\lambda$
and $\delta>0$ such that $F_\nu(z_1+z_2)<S-\delta$ for
$\lambda\le\nu\le\mu$. Add a third account of type $z_3$, with a position large enough that $F_\mu(z_3)>F_\lambda(z_3)$. Continuity gives a
level $\nu\in(\lambda,\mu]$ with
\[
 0<F_\nu(z_3)-F_\lambda(z_3)<\delta.
\]
Choose target $S+F_\lambda(z_3)$ for these three accounts. After merging
the first two, total residuals at $\nu$ are still below this target, so every
clearing level is greater than $\nu$. The third account receives strictly more;
by budget balance, the merged account receives strictly less. This is a
profitable split. If (M) fails, the same argument with a lower nearby
level gives a profitable merge. Binary operations suffice, since every
finite split or merge can be performed in stages.
\end{proof}

For finite rules, the lemma restricts the monotonicity of $b$ and the superadditivity of the appropriate signed reciprocal. For the boundary rules, it restricts the weight profile $w$.

\begin{proposition}[Splitting and merging]\label{prop:splitmerge}
Within the extended log-exponential family:

\begin{enumerate}
\item[(I)] The log-exponential rule is split-proof if and only if
\begin{enumerate}
\item[(S1)] $b$ is nondecreasing;
\item[(S2)] whenever $b(e_1+e_2)<0$,
\[
 \frac{1}{b(e_1+e_2)}
 \le
 \frac{1}{b(e_1)}+\frac{1}{b(e_2)}.
\]
\end{enumerate}

The rule is merge-proof if and only if
\begin{enumerate}
\item[(M1)] $b$ is nonincreasing;
\item[(M2)] whenever $b(e_1+e_2)>0$,
\[
 \frac{1}{b(e_1+e_2)}
 \ge
 \frac{1}{b(e_1)}+\frac{1}{b(e_2)}.
\]
\end{enumerate}

\item[(II)] Weighted CEA is split-proof if and only if $w$ is superadditive,
\[
 w(e_1+e_2)\ge w(e_1)+w(e_2),
\]
and merge-proof if and only if $w$ is nonincreasing.

\item[(III)] Weighted CEL is split-proof if and only if $w$ is nonincreasing, and merge-proof if and only if $w$ is superadditive.
\end{enumerate}

The proportional rule is the only member of the characterised class that is both split-proof and merge-proof.
\end{proposition}

The proof is in Appendix~\ref{app:splitting}.

\paragraph{Rules based only on position size.}
If the rule ignores equity, $b$ is constant in the finite case and $w$ is constant at either boundary. The boundary rules are then ordinary CEA and CEL. Proposition~\ref{prop:splitmerge} gives the following restrictions.

\begin{corollary}[Position-only rules]\label{cor:position-only}
A finite rule with constant $b$ is split-proof if and only if $b\ge0$, and merge-proof if and only if $b\le0$. Ordinary CEL is split-proof, ordinary CEA is merge-proof, and the proportional rule ($b=0$) is the only position-only member satisfying both properties.
\end{corollary}

For the finite homogeneous rules, the sign conditions are reversed: splitting divides equity together with the position, changing the parameter $b(x)=\theta p_\tau/x$. Combining Propositions~\ref{prop:homogeneous} and~\ref{prop:splitmerge} gives the full classification.

\begin{corollary}[Manipulation under homogeneity]\label{cor:homogeneous}
Among homogeneous members:
\begin{enumerate}
\item[(i)] the finite rule $r^\theta$ is split-proof if and only if $\theta\le0$, and merge-proof if and only if $\theta\ge0$;
\item[(ii)] $\mathrm{CEA}_{\gamma}$, with weight profile $w(x)=x^\gamma$, is split-proof if and only if $\gamma\ge1$, and merge-proof if and only if $\gamma\le0$;
\item[(iii)] $\mathrm{CEL}_{\gamma}$ is split-proof if and only if $\gamma\le0$, and merge-proof if and only if $\gamma\ge1$.
\end{enumerate}
Hence the split-proof homogeneous class consists of the finite family $\theta\le0$, the CEA boundaries $\gamma\ge1$, and the CEL boundaries $\gamma\le0$.
\end{corollary}

For $\theta<0$, $b(x)=\theta p_\tau/x$ is increasing and $1/(-b(x))$ is additive; $\theta>0$ gives the dual conditions. At the boundaries, $x^\gamma$ is superadditive exactly for $\gamma\ge1$ and nonincreasing exactly for $\gamma\le0$.

\section{Percentage reductions and fairness}\label{sec:fairness}
In this section, we consider how the burden of deleveraging is distributed across accounts.
We measure an account's burden by the fraction of its position
that is reduced. We compare these percentage reductions in the (position-weighted)
Lorenz order. Lorenz comparisons of claims rules have been studied systematically by
\citet{thomsonLorenz2012}. We look primarily at percentage reductions rather than absolute reductions, which are more standard in the claims literature. Percentage reductions naturally make proportional division fairest (Proposition~\ref{prop:prop-fairest}), whereas absolute reductions naturally make ordinary CEL fairest (Remark~\ref{rem:burden-measures}).

 For vectors $x,y\in\mathbb{R}^N$ and positive weights $w_i$, $i\in N$, we say that $x$ \textbf{$w$-Lorenz dominates} $y$ if for each $t\le\sum_i w_i$ the sum of the $t$ smallest coordinates of $x$ is at least the sum of the $t$ smallest coordinates of $y$, where coordinate $i$ is counted $w_i$ times; domination is strict if one of these inequalities is strict.\footnote{Neither $t$ nor the weights need be integers: a coordinate may be counted a fractional number of times. Formally, the sum of the $t$ smallest coordinates of $x$ is $\min\{\sum_i a_ix_i:0\le a_i\le w_i,\ \sum_i a_i=t\}$.} With unit weights, this is the usual Lorenz order. Throughout, we only compare vectors with the same weighted sum, $\sum_iw_ix_i=\sum_iw_iy_i$. Domination then means that $x$ is less dispersed than $y$.

 For a nonzero position profile, let $Q=\sum_iq_i$ and omit accounts
with $q_i=0$. The \textbf{percentage reduction} of account $i$ is
$h_i=1-r_i/q_i$. For two allocations $r$ and $s$ at the same problem $(q,e,\hat{E})$, we say that $r$ is
\textbf{at least as fair as} $s$ in the \textbf{weighted Lorenz order},
written $r\succeq_q^{\mathrm L}s$, if the percentage reductions under $r$ $q$-Lorenz dominate those under $s$. The comparison is strict if domination is strict.
For rules $r$ and $s$, write $r\succeq^{\mathrm L}s$ if
$r(q,e,\widehat E)\succeq_q^{\mathrm L}s(q,e,\widehat E)$
for every problem $(q,e,\widehat E)$.
Weighting by position in the Lorenz ranking makes sure that any two
allocations for the same problem have the same weighted mean
$\bar h=Q^{-1}\sum_{i\in N}q_ih_i=1-\widehat E/Q$.

The same comparison can be expressed through
\textbf{position-weighted convex order}. Since the weighted mean is fixed,
$r\succeq_q^{\mathrm L}s$ if and only if
\[
 \sum_iq_i\Phi\!\left(1-\frac{r_i}{q_i}\right)
 \le \sum_iq_i\Phi\!\left(1-\frac{s_i}{q_i}\right)
 \qquad\text{for every convex }\Phi:\R\to\R
\]
\citep[Section~14.B]{marshallOlkinArnold2011}. The fairness optimum is proportional division.

\begin{proposition}[Proportional division is fairest]\label{prop:prop-fairest}
For every nonzero position profile $q$ and feasible target $\widehat E$,
the proportional allocation is at least as fair as every feasible allocation:
\[
 r^{\mathrm{prop}}\succeq_q^{\mathrm L}r.
\]
The comparison is strict unless $r=r^{\mathrm{prop}}$.
\end{proposition}

\begin{proof}
Write $\alpha=\widehat E/Q$ and, for a convex function $\Phi$, let
$B_\Phi(r;q)=\sum_iq_i\Phi(1-r_i/q_i)$.
The weights sum to $Q$ and satisfy
\[
 \sum_iq_ih_i=Q-\widehat E=Q(1-\alpha).
\]
Jensen's inequality gives
$B_\Phi(r;q)\ge Q\Phi(1-\alpha)=B_\Phi(r^{\mathrm{prop}};q)$,
so $r^{\mathrm{prop}}\succeq_q^{\mathrm L}r$. Proportional division
attains it because every reduction equals $1-\alpha$; for strictly
convex $\Phi$, equality requires all reductions to equal their mean.
\end{proof}

In particular, proportional division is the unique fairness-optimal rule
under this percentage-reduction criterion within the class characterised
by Theorem~\ref{thm:type-comppair}.

\begin{remark}[Absolute reductions]\label{rem:burden-measures}
The fairest allocation depends on how burden is measured. If we look at absolute reductions $q_i-r_i$ rather than percentage reductions, ordinary CEL is fairest: its reductions Lorenz dominate those under every feasible allocation. This follows by duality from the Lorenz optimality of CEA~\citep{thomsonLorenz2012}, applied to reductions rather than residual positions. More generally, for a positive weight profile $w$, weighted CEL is fairest if we compare reductions per unit of weight $(q_i-r_i)/w(e_i)$ in the Lorenz order with weights $w(e_i)$. See also Remark~\ref{rem:absolute-reductions}.
\end{remark}

\subsection{Scaling the parameter profile}

Multiplying a finite parameter profile $b$ by a common factor $\kappa\ge0$ gives a family starting at proportional division, $\kappa=0$. The following theorem shows that percentage reductions become more dispersed as $\kappa$ increases.

\begin{theorem}[Fairness ranking]\label{thm:fairness-rays}
Fix a continuous parameter profile $b:\Rpos\to\R$, and let $r^\kappa$
denote the rule with parameter profile $b_\kappa(x)=\kappa b(x)$. Then
\[
 0\le\kappa_1\le\kappa_2<\infty
 \quad\Longrightarrow\quad
 r^{\kappa_1}\succeq^{\mathrm L}r^{\kappa_2}.
\]
\end{theorem}

The proof is in Appendix~\ref{app:fairness}.

\begin{corollary}[Fairness in the $\theta$ family]
\label{cor:theta-fairness}
The rules in the $\theta$ family satisfy
\begin{align*}
 -\infty\le\theta_1\le\theta_2\le0
 &\quad\Longrightarrow\quad
 r^{\theta_2}\succeq^{\mathrm L}r^{\theta_1},\\
 0\le\theta_1\le\theta_2\le+\infty
 &\quad\Longrightarrow\quad
 r^{\theta_1}\succeq^{\mathrm L}r^{\theta_2}.
\end{align*}
Thus, on either side of zero, moving $\theta$ towards zero improves
fairness, with proportional division at $\theta=0$.
\end{corollary}

\begin{proof}
For finite parameters, apply Theorem~\ref{thm:fairness-rays} to
$b(x)=-p_\tau/x$ with $\kappa=-\theta$ for the first comparison, and to
$b(x)=p_\tau/x$ with $\kappa=\theta$ for the second.
The endpoints follow from Remark~\ref{rem:type-endpoint-limits}
and continuity of the convex sums.
\end{proof}

\begin{remark}[Absolute reductions]\label{rem:absolute-reductions}
For absolute (instead of percentage) reductions per unit of equity, the comparison is monotone in
$\theta$ across the whole family. In particular, positive $\theta$ are fairer in this metric than the proportional rule and the negative $\theta$ rules.

This follows by duality from Proposition~\ref{prop:risk-monotonicity},
which we will prove later on: for
$-\infty\le\theta_1\le\theta_2\le+\infty$, the reductions
$((q_i-r_i^{\theta_2})/e_i)_{i\in N}$ $e$-Lorenz dominate
$((q_i-r_i^{\theta_1})/e_i)_{i\in N}$. For $b_\eta(x)=\eta/w(x)$ with $w>0$, the same argument shows that
reductions per unit of weight become less dispersed as $\eta$ increases,
using weights $w(e_i)$.

\end{remark}

\begin{remark}[Fairness at the homogeneous boundaries]\label{prop:boundary-comparison}
Write $\mathrm{CEA}_\gamma$ and $\mathrm{CEL}_\gamma$ for the boundary rules with
weight profile $w(x)=x^\gamma$.
 On the boundary, the percentage reduction comparison depends on the profile. For every $\gamma>1$, there are profiles on which $\mathrm{CEA}_\gamma$ is fairer than water-filling and profiles with the reverse comparison. The same holds for $\mathrm{CEL}_\gamma$ when $\gamma\le0$. Appendix~\ref{app:fairness} gives examples with strict weighted Lorenz dominance.
\end{remark}

\subsection{Universal fairness comparisons}

For finite rules, Theorem~\ref{thm:fairness-rays} gives universal fairness comparisons by common scaling. For non-positive profiles, this is also necessary.

\begin{proposition}[Universal fairness comparisons]
\label{prop:universal-fairness}
Let $b_1,b_2:\Rpos\to(-\infty,0]$ be continuous, and let $r^1,r^2$
denote the corresponding rules. Then $r^1\succeq^{\mathrm L}r^2$ if and only if
\[
 b_1=\kappa b_2\qquad\text{for some }0\le\kappa\le1.
\]
\end{proposition}

\begin{proof}
Sufficiency is Theorem~\ref{thm:fairness-rays}, including $\kappa=0$,
which gives proportional division.

For necessity, first suppose $b_2(e)=0$ at some equity $e$.
On a two-account profile with this common equity and distinct positive
positions, $r^2$ is proportional. At any interior target, the assumed
comparison and strict convexity force $r^1$ to give the same retention
fractions. For a common interior level,
$\varphi_b^{-1}(\lambda\varphi_b(q))/q$ is strictly increasing in $bq$.
Thus $b_1(e)q_1=b_1(e)q_2$, so $b_1(e)=0$.
In particular, if $b_2\equiv0$, then $b_1\equiv0$, and the conclusion
holds with $\kappa=0$.

Otherwise, take any $e_1,e_2>0$ with $b_2(e_i)<0$ and set
$q_i=-1/b_2(e_i)$. since $b_2(e_i)q_i=-1$, $r^2$ gives both accounts the same retention
fraction. This is the unique minimiser of
any strictly convex burden measure, so $r^1$ must give the same
fractions at every interior target. The same strict monotonicity gives
\[
 b_1(e_1)q_1=b_1(e_2)q_2,
 \qquad\text{or equivalently}\qquad
 \frac{b_1(e_1)}{b_2(e_1)}=\frac{b_1(e_2)}{b_2(e_2)}.
\]
The ratio is a common constant $\kappa\ge0$ wherever
$b_2<0$. Since $b_1$ also vanishes wherever $b_2=0$, we obtain
$b_1=\kappa b_2$ on the full domain.

If $\kappa>1$, choose an equity $e$ with $b_2(e)<0$ and two accounts
with this equity, distinct positive positions and an interior target.
Their scores $b_2(e)q_i$ differ. The derivative formula in the proof
of Theorem~\ref{thm:fairness-rays}, applied to the squared retention
fraction, is then strictly positive for $\kappa>0$. Scaling from $1$
to $\kappa$ increases squared burden dispersion as well, since its
mean is fixed. This contradicts the assumed fairness comparison.
Hence $\kappa\le1$.
\end{proof}

\section{Risk}\label{sec:risk}

Manipulation and fairness considerations both favour proportional division:
it is the only rule in our characterised class that is both split- and
merge-proof, and its percentage reductions Lorenz dominate those of every other allocation, while its total reductions Lorenz dominate those of every other split-proof rule in our main family. Other objectives can be in tension with this
choice. In particular, reducing platform risk may require larger
percentage reductions for more highly leveraged accounts.

Platform risk arises from losses on the positions that remain after ADL. In the
single-asset isolated-margin model of \citet{campbell}, aggregate
short-account loss at terminal price $p$ is
\begin{equation}\label{eq:campbell-loss}
 L(r,p)=\sum_i\max\{0,(p-p_\tau)r_i-e_i\},
\end{equation}
where $p_\tau$ is the market price at the ADL event and $e_i$ is equity valued at that price, as in Section~\ref{sec:model}. Execution at this price leaves equity unchanged. For $r_i>0$, uncovered loss begins when the price increase exceeds $e_i/r_i$, that is, when the relative price increase exceeds the reciprocal of surviving leverage $p_\tau r_i/e_i$. For a random terminal price $P$, platform risk is measured by a risk measure~\citep{Delbaen} $\rho(L(r,P))$. We assume that $\rho$ is monotone: a loss that is no greater in any state cannot have higher risk.

Recall the definition of weighted Lorenz dominance from the previous section. For feasible residual allocations $r,s$ with the same total, write
$r\preceq_e s$ if surviving leverage $(p_\tau r_i/e_i)_{i\in N}$ $e$-Lorenz dominates surviving leverage $(p_\tau s_i/e_i)_{i\in N}$. Equivalently, since the common factor $p_\tau$ does not affect the comparison,
\[
 \sum_i e_i\Phi(r_i/e_i)
 \le \sum_i e_i\Phi(s_i/e_i)
 \qquad\text{for every convex }\Phi:\R\to\R.
\]
 Thus $r\preceq_e s$ means that surviving leverage is less dispersed under $r$.
For rules $r$ and $s$, write $r\preceq s$ if
$r(q,e,\widehat E)\preceq_e s(q,e,\widehat E)$
for every problem $(q,e,\widehat E)$.
The next lemma identifies the allocation order with lower uncovered loss at every terminal price.

\begin{lemma}[Risk order]\label{lem:risk-order}
For every problem $(q,e\widehat E)$ and feasible residual allocations $r,s$ at the problem,
\[
 r\preceq_e s
 \quad\Longleftrightarrow\quad
 L(r,p)\le L(s,p)
 \qquad\text{for every terminal price }p\ge0.
\]
Consequently, $r\preceq_e s$ implies
\[
 \rho(L(r,P))\le\rho(L(s,P))
\]
for every random terminal price $P$ and every monotone risk measure $\rho$
for which the quantities are defined.
\end{lemma}

The proof is in Appendix~\ref{app:risk-details}.

\subsection{Universal risk optimality of water-filling}

Water-filling minimises risk over all feasible residual allocations.

\begin{proposition}[Water-filling is least risky \citep{campbell}]\label{prop:waterfill-risk}
For every problem $(q,e,\hat{E})$ the water-filling allocation $r^{\mathrm{WF}}$ at the problem satisfies
\[
 r^{\mathrm{WF}}\preceq_e s
\]
for every feasible residual allocation $s$ at the problem.
\end{proposition}

This is Campbell et al.'s state-by-state loss-minimisation result
\citeyearpar[Theorem~2.4(d)]{campbell}, expressed through
Lemma~\ref{lem:risk-order}. This comparison is over the full feasible allocation set and does not
require homogeneity or any of the axioms of
Theorem~\ref{thm:type-comppair}.

For a particular risk measure, other allocations may tie with water-filling. \textbf{CVaR (conditional value at risk)}, the average loss over a specified upper tail, can have a continuum of minimisers \citep[Theorem~2.16 and Appendix~A.5]{campbell}. This leaves a rule-level question: can another member of the class characterised in Theorem~\ref{thm:type-comppair} minimise the same fixed risk criterion on every problem? For \textbf{spectral risk measures}~\citep{acerbiSpectral2002}, represented below as mixtures of upper-tail CVaRs of losses, the following theorem gives uniqueness on our class under a full-support condition on prices.

\begin{theorem}[Unique universal risk minimisation]
\label{thm:risk-optimal-composition}
Fix an integrable terminal price $P$ with a density positive on
$(p_\tau,\infty)$ and a spectral risk measure
\[
 \rho(X)=\int_{[0,1)}\operatorname{CVaR}_\beta(X)\,\omega(d\beta),
 \qquad \operatorname{CVaR}_0(X)=\mathbb E[X],
\]
where $\omega$ is a probability measure and $\rho(L(r,P))$ is finite for
every feasible allocation. Within the full class of
Theorem~\ref{thm:type-comppair}, water-filling is the unique rule that
minimises this criterion on every problem.
\end{theorem}

The proof is in Appendix~\ref{app:risk-details}.

\subsection{Risk ranking within the $\theta$-family}

\begin{proposition}[Risk ranking]\label{prop:risk-monotonicity}
For $-\infty\le \theta_1\le \theta_2\le+\infty$,
\[
 r^{\theta_1}\preceq r^{\theta_2}.
\]
Hence $\theta\mapsto\rho(L(r^\theta(q,e,\widehat E),P))$ is weakly increasing
for every problem $(q,e,\widehat E)$ and every monotone risk measure for which it is defined.
\end{proposition}

The proof is in Appendix~\ref{app:risk-details}. Decreasing $\theta$ from zero towards $-\infty$ moves from proportional division towards water-filling. Every rule along this interval is split-proof, and uncovered loss weakly decreases at every terminal price.

\begin{corollary}[Non-positive parameters suffice]\label{cor:positive-dominated}
For every $0\le\theta\le+\infty$,
\[
 r^0\succeq^{\mathrm L}r^\theta,
 \qquad r^0\preceq r^\theta.
\]
Thus proportional division weakly dominates every positive parameter rule
in both percentage-reduction fairness and risk.
\end{corollary}

\begin{proof}
Combine Corollary~\ref{cor:theta-fairness} and
Proposition~\ref{prop:risk-monotonicity} with one parameter equal to zero.
\end{proof}

\begin{remark}[Other homogeneous boundary rules]\label{rem:homogeneous-boundary-risk}
The homogeneous class also contains the boundary rules with $w(x)=x^\gamma$ for $\gamma\ne1$, which lie outside the $\theta$-family. By Corollary~\ref{cor:homogeneous}, the additional split-proof members are $\mathrm{CEA}_\gamma$ with $\gamma>1$ and $\mathrm{CEL}_\gamma$ with $\gamma\le0$.

Water-filling is weakly less risky than all these boundary rules on every problem, by Proposition~\ref{prop:waterfill-risk}. Their risk comparison with any finite $r^\theta$ depends on the profile: for each boundary rule with $\gamma\ne1$ and each finite $\theta$, there are problems on which the boundary allocation is strictly lower in the risk order and problems with the reverse comparison. Appendix~\ref{app:boundary-risk} gives the comparisons. Thus neither rule universally risk-dominates the other.
\end{remark}

\subsection{Universal risk comparisons}

We now compare two finite rules uniformly across profiles. Proposition~\ref{prop:risk-monotonicity} supplies such a comparison when both profiles have the form $\theta p_\tau/x$. The next result shows that every nontrivial universal comparison is of this form.

\begin{proposition}[Universal risk comparisons]\label{prop:universal-risk}
Let $b_1,b_2:\Rpos\to\R$ be continuous parameter profiles,
and let $r^1,r^2$ denote the corresponding rules. Then
$r^1\preceq r^2$ if and only if either $b_1=b_2$, or
\[
 b_1(x)=\frac{\theta_1p_\tau}{x},\qquad b_2(x)=\frac{\theta_2p_\tau}{x}
 \quad\text{for constants }\theta_1\le \theta_2.
\]
\end{proposition}

The proof is in Appendix~\ref{app:risk-comparisons}.

\section{Fairness and risk}\label{sec:fairness-risk}

The fairness ranking in Theorem~\ref{thm:fairness-rays} and the risk
ranking in Proposition~\ref{prop:risk-monotonicity} run in opposite
directions within the $\theta$-family for
$\theta\le0$.

\begin{corollary}[Opposite Lorenz rankings]\label{cor:opposite-lorenz}
\[
 -\infty\le\theta_1\le\theta_2\le0
 \quad\Longrightarrow\quad
 r^{\theta_2}\succeq^{\mathrm L}r^{\theta_1}
 \quad\text{and}\quad
 r^{\theta_1}\preceq r^{\theta_2}.
\]
\end{corollary}

\begin{proof}
Combine Corollary~\ref{cor:theta-fairness} with
Proposition~\ref{prop:risk-monotonicity}.
\end{proof}

For every problem, moving $\theta$ from zero towards $-\infty$ weakly reduces uncovered loss at every terminal price and makes percentage reductions weakly more dispersed. Measuring the burden by reductions per unit of equity gives the same fairness ranking on this interval (Remark~\ref{rem:absolute-reductions}). For other homogeneous boundary rules this is not true:

\begin{remark}[Fairness and risk at the homogeneous boundaries]
For each homogeneous boundary rule outside the $\theta$-family
($\mathrm{CEA}_\gamma$ or $\mathrm{CEL}_\gamma$ with $\gamma\ne1$),
there is a problem on which  proportional division ($r^0$) is
strictly fairer and offers weakly lower platform risk.
Take two accounts with the same positive initial leverage and different
equities. For CEA, choose a small positive target so neither position
cap binds. For CEL, choose a small positive reduction so both residual
positions are positive. The weights per unit of equity,
$e_i^{\gamma-1}$, differ across accounts, so the boundary rule leaves
different surviving leverages and makes different percentage reductions.
Proportional division ($r^0$) gives both accounts the same percentage
reduction and leaves equal surviving leverages. The target fixes the
equity-weighted mean surviving leverage. By Jensen's inequality and
Lemma~\ref{lem:risk-order}, $r^0$ gives weakly lower uncovered loss at
every terminal price.
\end{remark}

The previous remark together with Corollary~\ref{cor:opposite-lorenz} can therefore justify to focus on the $\theta$-family with $-\infty\le\theta\le0$ when choosing a homogeneous path-independent rule through trading off platform risk and fairness.
\subsection{Calibration}
A platform can use historical deleveraging problems to choose a single $\theta$
parameter before applying the rule. Fix a nonempty set $\mathcal C$ of
calibration problems and a risk bound $\overline R$. For each
$\mathcal P=(q,e,\widehat E)\in\mathcal C$, let $R_{\mathcal P}$ be a
lower semicontinuous risk criterion on its feasible allocations such that
$r\preceq_e s$ implies $R_{\mathcal P}(r)\le R_{\mathcal P}(s)$.
For example, $R_{\mathcal P}(r)$ could be the uncovered loss at the
highest price observed over a fixed horizon after that market's ADL event,
as in Section~\ref{sec:empirical}.
The worst-case risk of $r^\theta$ on the calibration set is
\[
 R_{\mathcal C}(\theta)
 =\sup_{\mathcal P\in\mathcal C}R_{\mathcal P}\bigl(r^\theta(\mathcal P)\bigr).
\]
Thus $R_{\mathcal C}(\theta)\le\overline R$ requires the rule to meet
the risk bound at every calibration problem.

\begin{proposition}[Ex ante calibration of $\theta$]
\label{prop:homogeneous-selection}
The set $\Theta=\{\theta\in[-\infty,0]:R_{\mathcal C}(\theta)\le\overline R\}$
is a closed, possibly empty lower interval. If $\Theta$ is nonempty, then
$\theta^*=\max\Theta$ exists and $r^{\theta^*}\succeq^{\mathrm L}r^\theta$
for every $\theta\in\Theta$. The fixed rule $r^{\theta^*}$ satisfies
the axioms of Theorem~\ref{thm:type-comppair}, homogeneity and split-proofness.
\end{proposition}

\begin{proof}
For each $\mathcal P\in\mathcal C$, the map $\theta\mapsto r^\theta(\mathcal P)$
is continuous at finite parameters by the schedules and clearing, and at
$-\infty$ by Remark~\ref{rem:type-endpoint-limits}. Lower semicontinuity of
$R_{\mathcal P}$ therefore makes the set of parameters meeting the bound at
$\mathcal P$ closed. By Corollary~\ref{cor:opposite-lorenz}, risk is
nondecreasing in $\theta$ at each problem, so each such set is a lower
interval. Their intersection $\Theta$ is a closed lower interval in
$[-\infty,0]$ and hence has a greatest element whenever it is nonempty.
The same corollary gives the fairness comparison at every problem.
The axioms, homogeneity and split-proofness follow because $\theta^*$ is
fixed and nonpositive.
\end{proof}

Proportional division is selected if it meets the bound at every calibration
problem. If water-filling exceeds the bound at even one problem, no rule can
meet it there, by Proposition~\ref{prop:waterfill-risk}.

\subsection{Beyond homogeneity}

On a given profile, a non-homogeneous rule can attain minimum risk with greater fairness than every homogeneous rule that also attains minimum risk.

\begin{example}[Greater fairness at the same minimum CVaR]
\label{ex:heterogeneous-fairness}
Let $p_\tau=1$ and $P=p_\tau+D$ with $D$ exponentially distributed with mean one, and
take $\beta=1-e^{-2}$. The CVaR tail is $D\ge2$, corresponding to the threshold $r_i/e_i=1/2$. Take
\[
 e=(1,2,3),\qquad q=(3/4,4,9),\qquad \widehat E=15/2.
\]
Among feasible allocations, the CVaR-optimal set is given by $r_i\ge e_i/2$ and has minimum risk
$16.5$. Within this set, the fairest allocation in the weighted Lorenz order is
\[
 r^\dagger=\left(\frac12,\frac{28}{13},\frac{63}{13}\right),
 \qquad
 \left(\frac{r_i^\dagger}{q_i}\right)_i
 =\left(\frac23,\frac7{13},\frac7{13}\right).
\]
Consider the profile $b(x)=-\kappa_*/x^2$  where
$\kappa_*\approx5.30749$. This defines a path-independent, split-proof
rule on the full domain and selects this allocation at the displayed problem. Indeed, the second and
third accounts have the same $q_i/e_i^2$ and hence retain the same percentage. As $\kappa$ increases from zero, the first residual rises from $9/22$ to $15/28$ in the $\mathrm{CEA}_2$ limit, so it equals $1/2$ at a finite parameter.

No homogeneous rule selects $r^\dagger$, including
the additional CEA and CEL boundaries. Hence $r^\dagger$ is strictly
fairer in the weighted Lorenz order than every
homogeneous member attaining minimum CVaR.
The proof and the persistence of the advantage over the full
homogeneous class on an open neighbourhood are in
Appendix~\ref{app:heterogeneous-selection}.
\end{example}

\section{Counterfactual replay}\label{sec:empirical}

We illustrate the fairness--risk trade-off using the Hyperliquid ADL
episode of 10 October 2025. We compare the observed reductions with
allocations from $r^\theta$ at the same market-level targets. The
underlying data come from ConejoCapital's
HyperMultiAssetedADL~\citep{conejoHyperMultiAssetedADL2025}
and HyperReplay~\citep{conejoHyperReplay2025}
repositories, which provide ADL fills, market data and clearinghouse
snapshots from Hyperliquid's public archive, with support from
Hydromancer. We reconstruct the full population before each selected
batch and evaluate alternative allocations on this population.

\subsection{Replay design and valuation}

For each of 18 markets with price coverage, we take the first ADL batch
accounting for at least 1\% of that market's recorded ADL volume. These
batches occur between 21:16 and 21:19 UTC, on an approximately 3 minute long time interval. Positions and cash balances
are reconstructed immediately before the relevant batch from a
clearinghouse snapshot and subsequent fills, funding and ledger events.

Hyperliquid closes auto-deleveraged positions at the previous mark
price~\citep{hyperliquid2026}, so the volume-weighted execution price
$p_{\tau_c}$ of market $c$'s batch is the exchange's mark price for that
market at the batch time. The mark price updates approximately every three
seconds, whenever validators publish new oracle prices~\citep{hyperliquidMarkPrice}.
The mark lagged Hyperliquid's own trades during the crash: the
execution price exceeded the last trade before the batch by up to 11\%, in 16 of the 18 markets. We value every position of every
account at the mark price at the batch time: the market's own batch
price where an ADL fill exists in that block, otherwise a reconstructed
mark.\footnote{Only one-minute marks are published. We rebuild the
three-second mark from its documented formula~\citep{hyperliquidMarkPrice}
using Hyperliquid's archived order books and trades and Binance prices,
pinned to Hyperliquid's archived one-minute marks and the batch prices. Against the
batch prices, its median error is 0.9\%, half that of interpolated
one-minute marks. Details are in the reproduction files.}
Account equity is cash plus unrealised P\&L
at these marks, and is allocated across the account's positions in
proportion to their marked notionals. As in Section~\ref{sec:model}, positions $q_{ic}$ and residuals $r_{ic}$ in
market $c$ are in units of asset $c$, and equities $e_{ic}$ are in
dollars at $p_{\tau_c}$. In market $c$, the rule $r^\theta$ has profile
$b(x)=\theta p_{\tau_c}/x$, so that $\theta$ multiplies the leverage
$p_{\tau_c}q_{ic}/e_{ic}$. The same $\theta$ is used in every market.

We exclude positions of accounts created after the snapshot, whose cash
is unknown, 23 positions with non-positive marked equity, which lie
outside the model's domain, and 32 positions whose leverage exceeds
twice the market's maximum, which are replay artifacts. Exclusions apply
to every rule, the observed benchmark, all denominators and the market
targets. The resulting sample contains 39,915 account--market
observations with initial short notional of \$5.219 billion. The
observed reductions remove \$534.588 million, or 10.24\% of that
notional, and cover 99.79\% of ADL notional in the selected batches.
Each counterfactual leaves the same surviving notional in each market.
The exercise reallocates these first batches; it does not replay the
entire subsequent ADL cascade.

The observed allocation fully closes 12,333 of the 12,378 positions it
reduces in this sample. Hyperliquid's documentation
describes a ranking based on unrealised P\&L and leverage~\citep{hyperliquid2026},
linking this benchmark to the entry-price dependence mentioned in
Section~\ref{sec:manipulation}. We use the recorded reductions without
reconstructing the ranking algorithm.

\subsection{Risk and fairness statistics}

We hold residual positions passively and evaluate them at the batch time and along the subsequent price path. Prices are the reconstructed Hyperliquid marks from the batch time through 17 October. Let $\mathrm{mm}_c$ denote the \textbf{maintenance-margin rate} in market $c$, i.e. the minimum equity required as a fraction of the position's current notional value. For a short position, the price at which equity falls to this requirement is
\[
 p^*_{ic}=\frac{e_{ic}+p_{\tau_c}r_{ic}}
 {r_{ic}(1+\mathrm{mm}_c)},\qquad r_{ic}>0.
\]
We report the share of surviving notional whose threshold is reached
by each horizon, aggregating with surviving notional as weights.
These are threshold breaches of passive positions, not observed
subsequent liquidations.

To connect the exercise to Section~\ref{sec:risk}, define
\[
 L^{\mathrm{peak}}(r,\Delta)
 =\sum_{c,i}\left[
 \left(P_c^{\max}(\Delta)-p_{\tau_c}\right)r_{ic}-e_{ic}
 \right]_+,
\]
where $P_c^{\max}(\Delta)$ is the maximum price from the batch time
through horizon $\Delta$. This evaluates the loss~\eqref{eq:campbell-loss}
of each market at its own peak price and sums over markets; the peaks
need not coincide. It is a stress criterion for the passive
residual positions, not a realised loss of the platform or its insurance
fund. Because large shortfalls can remain under every allocation, we report
the \textbf{excess peak loss}
\[
 L^{\mathrm{excess}}(r,\Delta)
 =L^{\mathrm{peak}}(r,\Delta)
  -L^{\mathrm{peak}}(r^{-\infty},\Delta).
\]
Subtracting the same water-filling benchmark leaves rankings unchanged.
Proposition~\ref{prop:risk-monotonicity} orders both loss statistics
within $r^\theta$. It does not order maintenance-margin breach shares,
which depend on the amount of notional near the threshold.

Within market $c$, write $Q_c=\sum_iq_{ic}$,
$h_{ic}=1-r_{ic}/q_{ic}$ and $\bar h_c=Q_c^{-1}\sum_iq_{ic}h_{ic}$.
We report the \textbf{position-weighted Gini coefficient} of percentage reductions (the \textbf{burden Gini}),
\[
 G_c=\frac{\sum_{i,j}q_{ic}q_{jc}|h_{ic}-h_{jc}|}{2Q_c^2\bar h_c},
\]
their \textbf{mean absolute deviation (MAD)},
$Q_c^{-1}\sum_iq_{ic}|h_{ic}-\bar h_c|$, and the share of reductions borne by
the most heavily reduced 10\% of initial position notional. Under
proportional division, Gini and MAD are zero and this top-decile share
is 10\%. All three statistics respect the weighted Lorenz order, so
Corollary~\ref{cor:theta-fairness} orders them within each market.
The \textbf{weighted variance} is $Q_c^{-1}\sum_iq_{ic}(h_{ic}-\bar h_c)^2$.
Dividing the convex sum $\sum_iq_{ic}\Phi(h_{ic})$ by $Q_c$ gives MAD
for $\Phi(h)=|h-\bar h_c|$, and variance plus the fixed squared mean
$\bar h_c^2$ for $\Phi(h)=h^2$.
We average the market statistics with weights $p_{\tau_c}Q_c$, the
markets' initial notionals, preserving these comparisons. MAD is expressed in percentage points. Comparing within
markets avoids treating differences in their required aggregate
reductions as unfairness.

We also compare absolute reductions per unit of equity,
$z_{ic}=p_{\tau_c}(q_{ic}-r_{ic})/e_{ic}$, the reduction in dollar
leverage. We weight accounts by $e_{ic}$.
The Gini and MAD are defined as above, replacing $h_{ic}$, $q_{ic}$ and
$Q_c$ by $z_{ic}$, $e_{ic}$ and $\sum_j e_{jc}$, respectively,
and using the equity-weighted mean of $z_{ic}$.
We again average market statistics using initial-notional weights.

\subsection{Results}

\begin{figure}[!t]
\centering
\includegraphics[width=\textwidth]{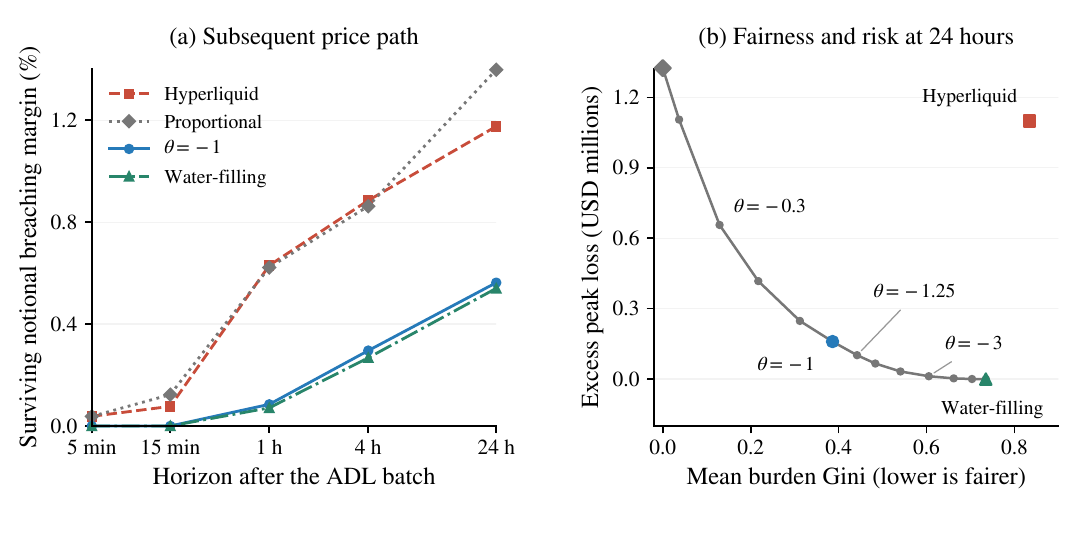}
\caption{Risk and fairness in the 10 October 2025 replay, using equity
marked at the batch time. Panel (a) follows passive
residual positions along the subsequent price path. Panel (b) compares
burden concentration with excess peak loss at 24 hours; dots show the
evaluated non-positive parameters and the water-filling limit. Lines
connect evaluated points. Both axes in panel (b) favour smaller values.}
\label{fig:replay}
\end{figure}

Every evaluated $\theta\le-0.5$, including water-filling, improves on the observed
allocation in every statistic in Table~\ref{tab:replay}.
Figure~\ref{fig:replay} shows the short-horizon breach shares and the fairness--risk trade-off across the evaluated grid. Water-filling itself
has more equal percentage reductions than the observed allocation on
all three fairness statistics. Intermediate parameters improve fairness
further while retaining lower risk than the observed allocation.
The observed allocation's high burden Gini reflects the concentration
of reductions into almost complete closures for affected accounts,
while unaffected accounts bear no reduction.

\begin{table}[!t]
\centering
\small
\setlength{\tabcolsep}{4pt}
\begin{tabular}{@{}lrrrrrrr@{}}
\hline
& \multicolumn{3}{c}{Burden dispersion}
& \multicolumn{3}{c}{Notional breaching margin (\%)}
& \shortstack{Excess loss\\(\$1,000)} \\
Rule & Gini & MAD & Top 10\% & 1 hour & 4 hours & 24 hours & 24 hours \\
\hline
Hyperliquid & 0.834 & 15.42 & 65.5 & 0.630 & 0.885 & 1.176 & 1,099.3 \\
Water-filling & 0.735 & 10.43 & 54.9 & 0.071 & 0.268 & 0.540 & 0.0 \\
$\theta=-3$ & 0.605 & 7.93 & 46.3 & 0.071 & 0.277 & 0.541 & 12.0 \\
$\theta=-1.25$ & 0.442 & 4.78 & 36.4 & 0.083 & 0.289 & 0.558 & 101.7 \\
$\theta=-1$ & 0.386 & 3.99 & 32.8 & 0.085 & 0.296 & 0.563 & 159.7 \\
$\theta=-0.5$ & 0.217 & 2.07 & 22.1 & 0.192 & 0.435 & 0.970 & 417.3 \\
$\theta=-0.3$ & 0.129 & 1.23 & 16.8 & 0.233 & 0.579 & 1.113 & 656.6 \\
Proportional & 0.000 & 0.00 & 10.0 & 0.622 & 0.863 & 1.398 & 1,324.8 \\
\hline
\end{tabular}

\caption{Allocations using equity marked at the batch time in 18 markets.
Fairness statistics are market averages weighted by initial notional.
MAD is in percentage points and the top-decile share is in percent.
Breach shares use surviving notional as weights. Excess loss is the
sum of market-specific peak uncovered losses above water-filling;
the latter has total peak loss of \$3.579 million at 24 hours.}
\label{tab:replay}
\end{table}

For example, $\theta=-1$ gives a 24-hour breach share of 0.563\%,
compared with 1.176\% under the observed allocation. Its excess peak
loss is \$159,684, compared with \$1,099,349 under the observed allocation.
Its burden Gini is 0.386 and its MAD is 3.99 percentage points,
compared with 0.735 and 10.43 under water-filling. Thus an intermediate
rule can distribute reductions substantially more equally than
water-filling while retaining most of its loss reduction relative to
the observed allocation.

Table~\ref{tab:equity-reductions} reports dispersion of reductions per
unit of equity. Both statistics decrease with $\theta$, including
for positive parameters, as in Remark~\ref{rem:absolute-reductions}. The results corroborate that an intermediate
rule can distribute reductions substantially more equally also if fairness is benchmarked in absolute reductions rather than percentage reductions.

\begin{table}[!t]
\centering
\small
\setlength{\tabcolsep}{7pt}
\begin{tabular}{@{}lrr@{}}
\hline
Rule & Gini & MAD \\
\hline
Hyperliquid & 0.8941 & 0.2956 \\
Water-filling & 0.8575 & 0.2557 \\
$\theta=-3$ & 0.7657 & 0.2075 \\
$\theta=-1.25$ & 0.6410 & 0.1583 \\
$\theta=-1$ & 0.5982 & 0.1453 \\
$\theta=-0.5$ & 0.4666 & 0.1122 \\
$\theta=-0.3$ & 0.3966 & 0.0967 \\
Proportional & 0.2908 & 0.0745 \\
$\theta=0.3$ & 0.2224 & 0.0584 \\
$\theta=1$ & 0.1285 & 0.0371 \\
$\theta=3$ & 0.0502 & 0.0186 \\
$\theta=10$ & 0.0176 & 0.0076 \\
Equity-weighted CEL & 0.0087 & 0.0045 \\
\hline
\end{tabular}

\caption{Dispersion of dollar reductions per unit of equity in the same
18 markets. Accounts are weighted by equity within each market;
market statistics are averaged using initial-notional weights.
MAD is in units of dollar leverage.
The corresponding mean is 0.1831 under every rule.}
\label{tab:equity-reductions}
\end{table}

The choice among intermediate parameters depends on the statistic.
At $\theta=-1.25$, the 24-hour breach share is 0.558\%, slightly below
the 0.563\% at $\theta=-1$, while the Gini rises from 0.386 to 0.442 and
the excess peak loss falls from \$159,684 to \$101,729.

The total peak-loss level is dominated by Zcash (ZEC) and Worldcoin (WLD) market accounts. At 24 hours,
their breach shares are approximately 39--40\% and 49--52\% across
the evaluated allocations. Under water-filling they account for
\$3.573 million of the \$3.579 million peak loss. Reporting excess loss
separates this common shortfall from the differences between allocations.
At three days, the observed breach share is 3.67\%, exceeding even
proportional division's 2.44\%. The full horizon and parameter results
are included with the reproduction code.

\subsection{Sensitivity and interpretation}

We also evaluate the allocations on 300 bootstrap price paths
constructed from 240-minute blocks of Binance minute returns, using
either the preceding 30 days or the subsequent week. At 24 hours,
the expected breach share for $\theta=-1$ is 0.174\% versus 0.270\%
for the observed allocation using the earlier sample, and 0.253\%
versus 0.451\% using the later sample.
Valuation matters for comparisons between nearby parameters but not for
the rankings. Valuing every position at the last trade price seen in the
replay instead of the mark, which overstates the equity of shorts in the
coin being deleveraged and understates it in their other positions,
gives 24-hour breach shares of 0.564\% at $\theta=-1$ and 1.232\% under
the observed allocation, and marking only the deleveraged coin gives
0.805\% and 1.334\%. The higher share at $\theta=-1$ under the latter
comes from a single large position in the Ethena (ENA) market whose
threshold lies close to the 24-hour peak. The reproduction files retain these valuations as
diagnostics.

Account equity is
allocated across positions in proportion to their marked notionals, and
each position is then evaluated separately. This does not reproduce
cross-margin liquidation. Marks without an ADL fill in the block are
reconstructed rather than observed.
Maintenance-margin rates are fixed at the replay's recorded metadata values,
without reconstructing historical margin tiers. The passive evaluation
omits subsequent trading, collateral transfers, funding, liquidations
and price impact. These limitations become more consequential at longer
horizons.

\section{Conclusion}

Path independence restricts ADL design to the extended log-exponential family. Proportional division is the only member that is both split- and merge-proof, and reopening-proofness requires entry-price independence. With fairness measured by dispersion of percentage reductions, proportional division weakly dominates positive parameters in both fairness and risk. On $-\infty\le\theta\le0$, moving from proportional division towards water-filling lowers risk at the cost of less equal percentage reductions. Heterogeneous rules can sometimes attain the same risk with greater fairness. The 10 October replay illustrates the size of this trade-off: several intermediate parameters improve on the observed allocation in both fairness and risk statistics, while distributing percentage reductions more equally than water-filling. The replay evaluates one stress episode under passive-position and single-asset assumptions.

\bibliographystyle{plainnat}
\bibliography{references}

\clearpage
\appendix

\section{Proof of Theorem~\ref{thm:type-comppair}}\label{app:characterisation}

A rule is \textbf{type-parametric} if there are an interval $\Lambda$
and a jointly continuous schedule
$f:(\R_+\times\Rpos)\times\Lambda\to\R_+$ such that, for each
type $(\bar q,\bar e)$, the schedule $\lambda\mapsto f((\bar q,\bar e),\lambda)$ is
nondecreasing from $0$ to $\bar q$, and each allocation is obtained at a common
level $\lambda\geq0 $ satisfying $\sum_i f((q_i,e_i),\lambda)=\widehat E$.

By Young's representation theorem \citeyearpar{young1987}, extended to separable type spaces by \citet[Theorem~1]{kaminski2006}, equal treatment of equal accounts, continuity and consistency give a parametric representation.

\begin{proposition}[Young representation \citep{kaminski2006}]\label{prop:young}
A rule satisfies equal treatment of equal accounts, continuity and consistency if and only if it admits a continuous type-parametric representation. The parameter interval can be chosen as $[0,1]$.
\end{proposition}

\begin{proof}
For positive positions, apply \citet[Theorem~1]{kaminski2006}
to the separable type space $(0,\infty)^2$ with cap $(\bar q,\bar e)\mapsto \bar q$.
His symmetry axiom is equal treatment of equal accounts, and our
consistency implies bilateral consistency. Normalise the resulting
continuous representation to $[0,1]$. Extending it by
$f((0,\bar e),\lambda)=0$ preserves joint continuity because
$0\le f((\bar q,\bar e),\lambda)\le \bar q$; consistency permits zero-position
accounts to be removed.

Conversely, take any continuous representation on $[0,1]$, adjoining
endpoint limits if needed. Monotonicity ensures that all clearing
levels give the same allocation. The common schedule gives equal
treatment, and the same level clears each reduced problem, giving
consistency. For convergent problems, take any convergent subsequence
of their clearing levels. Joint continuity makes the limiting level
clear the limiting problem. Uniqueness of its allocation proves
continuity of the rule.
\end{proof}

Fix a rule satisfying the axioms, and take the representation from
Proposition~\ref{prop:young}. Following \citet[Section~2.5.2]{schlegel}, write
\[
 T^{\bar e}_\lambda(\bar q)=f((\bar q,\bar e),\lambda),\qquad
 \mathcal L^{\bar e}_\lambda(\bar q)=\bar q-T^{\bar e}_\lambda(\bar q),\qquad 0\le\lambda\le1.
\]
Here $T^{\bar e}_\lambda$ is the \textbf{gain (residual-position) map} and $\mathcal L^{\bar e}_\lambda$
the \textbf{loss (buyback) map} at equity $\bar e$.
Gain maps, with their common parametrisation, fully determine a rule.
We first apply the one-dimensional characterisation of
\citet[Theorem~2.2]{schlegel} to profiles with a common equity.
For each equity, this gives a canonical scalar Young representation.
It remains to show that these scalar representations can be used
together to represent the original rule on arbitrary equity profiles.

Lemma~\ref{lem:common-levels} uses DPI to obtain composition
identities valid simultaneously at all equities; applying it to the
dual rule gives the corresponding identities under UPI.
Lemma~\ref{lem:no-mixed-cases} rules out different boundary cases at
different common equities.
Using the composition identities, Lemma~\ref{lem:finite-sync} shows
that the canonical finite schedules jointly represent the original
rule, and Lemma~\ref{lem:cea-sync} identifies the equity-dependent
weights in the boundary schedules.

We use duality throughout: the \textbf{dual rule} is
$r^*(q,e,E)=q-r(q,e,\sum_iq_i-E)$. It interchanges gain and
loss maps, DPI and UPI, and CEA and CEL; it sends a finite profile
$b$ to $-b$. The other axioms are preserved.

\begin{lemma}[Composition]\label{lem:common-levels}
For every $\lambda,\mu\in[0,1]$, the following statements hold.
\begin{enumerate}
\item[(a)] There exists $\nu\in[0,1]$ such that
\[
 T^{\bar e}_\nu=T^{\bar e}_\mu\circ T^{\bar e}_\lambda\qquad(\bar e>0).
\]
\item[(b)] If $\lambda\le\mu$, there exists $\nu\in[0,1]$ such that
\begin{equation}\label{eq:common-quotient}
 T^{\bar e}_\lambda=T^{\bar e}_\nu\circ T^{\bar e}_\mu\qquad(\bar e>0).
\end{equation}
\end{enumerate}
\end{lemma}

\begin{proof}
Part (a) follows from the proof of
\citet[Lemma~2.11]{schlegel}, with its intervals indexed by
types $(\bar q,\bar e)$ rather than claims; equity remains fixed under composition.
For \eqref{eq:common-quotient}, DPI between levels $\lambda\le\mu$
gives a common $\nu$ on each pair of types. The admissible levels for
each type form a nonempty closed interval in $[0,1]$, so the same
interval argument gives one $\nu$ valid for all types.
\end{proof}

\begin{lemma}[Behaviour on equal-equity profiles]\label{lem:no-mixed-cases}
For each $\bar e>0$, there is a member of the scalar extended
log-exponential family with which $r$ agrees at every profile satisfying
$e_i=\bar e$ for all $i$, for every target.
If $r$ agrees with CEA on all such profiles and targets for some
$\bar e>0$, then it agrees with CEA on every equal-equity profile
and target; the same implication holds for CEL.
Otherwise, for each common equity $\bar e$, these allocations are given
by a finite log-exponential parameter $b(\bar e)$.
\end{lemma}

\begin{proof}
The first assertion follows from \citet[Theorem~2.2]{schlegel}.
It remains to show that agreement with CEA or CEL on profiles with
one common equity implies the same agreement on all equal-equity profiles. The scalar characterisation identifies the gain maps at each equity level $\bar e>0$,
up to a relabelling of the parameter.

Call a Young level $\lambda$ binary if
$T^{\bar e}_\lambda\in\{0,\id\}$ for every $\bar e>0$.
At such a level, continuity in equity implies that either
$T^{\bar e}_\lambda=0$ for every $\bar e$ or
$T^{\bar e}_\lambda=\id$ for every $\bar e$.

Starting from any level $\lambda$, iterate its gain map, then
iterate the loss map complementary to the first limit. Each limit belongs
to the appropriate global family: Lemma~\ref{lem:common-levels}(a) and duality supply levels
for the iterates, and a convergent subsequence of levels represents
the pointwise limit. The gain map complementary to the second limit is
binary. If the maps at equity $\bar e$ have the finite log-exponential
or CEL form, it is the identity exactly when $T^{\bar e}_\lambda$ was
already the identity. In the CEA case, write
\[
 T^{\bar e}_\lambda(\bar q)=\min\{\bar q,A_{\bar e}(\lambda)\}
\]
for a continuous, nondecreasing
and onto function
$A_{\bar e}:[0,1]\to[0,\infty]$.
If they are cap maps, it is the identity exactly
when $A_{\bar e}(\lambda)>0$: cap maps are idempotent, and
iteration of their buybacks gives $(\bar q-nA_{\bar e}(\lambda))_+\to0$ when
the cap is positive.

Suppose $r$ agrees with CEA on all profiles with common equity $e_*$,
for every target, and let
$\alpha=\max\{\lambda:A_{e_*}(\lambda)=0\}$. For every
$\lambda>\alpha$, the binary limit is the identity at $e_*$, hence
everywhere. At any equity $\bar e$ where the gain maps are not cap maps,
we must therefore already have
$T^{\bar e}_\lambda=\id$. Continuity gives $T^{\bar e}_\alpha=\id$ there, whereas
$T^{e_*}_\alpha=0$. The binary limit at $\alpha$ would then take
both values, a contradiction. Thus the gain maps are cap maps at every
equity, so $r$ agrees with CEA on every equal-equity profile and target.
Applying this conclusion to the dual rule proves the assertion for CEL.
\end{proof}

In the finite case of Lemma~\ref{lem:no-mixed-cases}, the canonical
Young representation at common equity $\bar e$ is
\[
 g_{\bar e}(\bar q,u)
 =\varphi_{b(\bar e)}^{-1}
   (u\varphi_{b(\bar e)}(\bar q)),\qquad 0\le u\le1.
\]
The multi-dimensional Young representation $f((\bar q,\bar e),\lambda)$ and
$g_{\bar e}(\bar q,u)$ represent the same rule on profiles with common
equity $\bar e$. This does not yet identify their parameters:
the conversion from the Young level $\lambda$ to the canonical
multiplier $u$ may depend on $\bar e$.
The next lemma uses DPI and UPI to show that one common conversion
works for every equity. Consequently, the canonical schedules
represent the original rule also on unequal-equity profiles.

In the CEA case, $A_{\bar e}(\lambda)$ from the preceding proof is
the canonical cap corresponding to the Young level $\lambda$.
Lemma~\ref{lem:cea-sync} shows that these caps differ only by positive
equity-dependent factors; the CEL case follows by duality.

\begin{lemma}[Canonical representation in the finite case]\label{lem:finite-sync}
Suppose that, for each $\bar e>0$, $r$ agrees on all profiles with common
equity $\bar e$, for every target, with a finite log-exponential rule
with parameter $b(\bar e)$.
Then $b$ is continuous and the schedule
$\varphi_{b(\bar e)}^{-1}
   (u\varphi_{b(\bar e)}(\bar q))$ is a Young
representation of $r$ on the full domain.
\end{lemma}

\begin{proof}
For each $\bar e$, agreement on equal-equity profiles gives a
continuous, nondecreasing surjection $\chi_{\bar e}:[0,1]\to[0,1]$
such that
\begin{equation}\label{eq:scalar-reparam}
 f((\bar q,\bar e),\lambda)
 =T^{\bar e}_\lambda(\bar q)
 =g_{\bar e}(\bar q,\chi_{\bar e}(\lambda)).
\end{equation}
To see this, fix a positive reference position. Its residual under
$f$ determines a unique canonical multiplier, since $g_{\bar e}$
is strictly increasing in $u$ at that position. Considering the
reference position together with any other position shows that
the same multiplier gives both residuals. Continuity, monotonicity
and the endpoint values of $f$ give the stated properties of
$\chi_{\bar e}$.
We must show that $\chi_{\bar e}$ is independent of equity; this will
allow a single change of parameter for the full rule.

Fix a reference equity $e_*>0$. The endpoint levels are shared across
equities. Indeed, iterating $T^{\bar e}_\lambda$ multiplies its canonical
multiplier repeatedly, so the limit is $\id$ when
$\chi_{\bar e}(\lambda)=1$ and zero otherwise.
Lemma~\ref{lem:common-levels}(a) and compactness of $[0,1]$ represent
these limits at one common Young level. This level is binary, hence
has the same map at every equity, as in the proof of
Lemma~\ref{lem:no-mixed-cases}.
Thus $\chi_{e_*}(\lambda)=1$ implies
$\chi_{\bar e}(\lambda)=1$ for every $\bar e$.
Similarly, iterating the loss maps and using
Lemma~\ref{lem:common-levels}(a) by duality gives the identity exactly at equities
where $\chi_{\bar e}(\lambda)=0$. The complementary gain map is binary,
so $\chi_{e_*}(\lambda)=0$ implies
$\chi_{\bar e}(\lambda)=0$ for every $\bar e$.

The functions $\chi_{\bar e}$ also have the same level sets.
Suppose $\lambda\le\mu$ and
$\chi_{e_*}(\lambda)=\chi_{e_*}(\mu)=c<1$.
By the dual of \eqref{eq:common-quotient} and the scalar loss-map formula,
\[
 1-\chi_{\bar e}(\mu)
 =(1-\chi_{\bar e}(\nu))(1-\chi_{\bar e}(\lambda)).
\]
At $e_*$ this gives $\chi_{e_*}(\nu)=0$, hence
$\chi_{\bar e}(\nu)=0$ everywhere. Therefore
$\chi_{\bar e}(\lambda)=\chi_{\bar e}(\mu)$.
The case $c=1$ follows from the shared endpoints.
Interchanging reference equities gives the converse.
Consequently, there is an increasing homeomorphism
$h_{\bar e}:[0,1]\to[0,1]$ satisfying
\[
 \chi_{\bar e}=h_{\bar e}\circ\chi_{e_*}.
\]

We now use the two path-independence axioms to determine $h_{\bar e}$.
Choose Young levels $\lambda,\mu$ with
$\chi_{e_*}(\lambda)=u$ and $\chi_{e_*}(\mu)=v$.
By DPI, Lemma~\ref{lem:common-levels}(a) gives a common level $\nu$
for the composition of their gain maps.
Canonical gain maps compose by multiplying their multipliers, so
\[
 \chi_{e_*}(\nu)=uv,\qquad
 \chi_{\bar e}(\nu)=\chi_{\bar e}(\lambda)\chi_{\bar e}(\mu).
\]
It follows that
\[
 h_{\bar e}(uv)=h_{\bar e}(u)h_{\bar e}(v).
\]
By UPI, applying Lemma~\ref{lem:common-levels}(a) to the dual rule
gives a common level for loss-map composition. The canonical loss map is
\[
 \bar q-g_{\bar e}(\bar q,u)
 =\varphi_{-b(\bar e)}^{-1}
   ((1-u)\varphi_{-b(\bar e)}(\bar q)),
\]
so composing loss maps at multipliers $u,v$ gives the loss map at
$u+v-uv$. Hence
\[
 h_{\bar e}(u+v-uv)
 =h_{\bar e}(u)+h_{\bar e}(v)-h_{\bar e}(u)h_{\bar e}(v).
\]
The gain identity and continuity give
$h_{\bar e}(u)=u^{\zeta_{\bar e}}$ for some $\zeta_{\bar e}>0$:
the function $t\mapsto-\log h_{\bar e}(e^{-t})$ is continuous and additive.
In the loss identity, set $v=u>0$ and divide by $u^{\zeta_{\bar e}}$:
\[
 (2-u)^{\zeta_{\bar e}}=2-u^{\zeta_{\bar e}}.
\]
Letting $u\downarrow0$ gives $\zeta_{\bar e}=1$.
Thus $h_{\bar e}$ is the identity and all $\chi_{\bar e}$ coincide.

Write $\chi$ for their common value. For any profile $(q,e)$, including
unequal equities, the original representation therefore gives
\[
 r_i(q,e,\widehat E)=g_{e_i}(q_i,u),\qquad
 \sum_i g_{e_i}(q_i,u)=\widehat E,\qquad u=\chi(\lambda).
\]
Since $\chi$ is onto, the canonical schedule represents the original
rule on the full domain.

Finally, fix $\bar q>0$ and choose a Young level $\lambda$ with
$u=\chi(\lambda)\in(0,1)$. For this fixed $u$, the map
\[
 b\longmapsto\varphi_b^{-1}(u\varphi_b(\bar q))
\]
is continuous and strictly increasing: after division by $\bar q$,
it is the secant slope from zero of the strictly convex function
$z\mapsto\log(1-u+u e^z)$.
Its inverse is continuous. Applying this inverse to the jointly
continuous schedule $f((\bar q,\bar e),\lambda)$ in
\eqref{eq:scalar-reparam} recovers $b(\bar e)$.
Thus $b$ is continuous, as is the canonical schedule.
\end{proof}

\begin{lemma}[Proportional CEA thresholds]\label{lem:cea-sync}
If $r$ agrees with CEA on every equal-equity profile and target, then
$A_{\bar e}(\lambda)=w(\bar e)A_{e_*}(\lambda)$ for a positive continuous
function $w$ and any fixed reference equity $e_*$.
\end{lemma}

\begin{proof}
If $A_{e_*}(\lambda)=0$, the gain map complementary to the limit of the iterates of the loss maps $\mathcal L^{\bar e}_\lambda$ is binary. It is zero at $e_*$ and the identity at any equity with a positive threshold. Hence $A_{\bar e}(\lambda)=0$ everywhere.

If $\lambda\le\mu$ and both reference thresholds equal a finite
value, the dual of \eqref{eq:common-quotient} gives
\[
 A_{\bar e}(\mu)=A_{\bar e}(\lambda)+A_{\bar e}(\nu).
\]
At $e_*$, $A_{e_*}(\nu)=0$, so all thresholds at $\nu$ vanish.
Therefore the common value at $e_*$ determines the threshold at every
equity. Write $h_{\bar e}(t)=A_{\bar e}(\lambda)$ when $A_{e_*}(\lambda)=t<\infty$.
Loss-map composition gives
\[
 h_{\bar e}(s+t)=h_{\bar e}(s)+h_{\bar e}(t),\qquad s,t\ge0.
\]
These values are finite. Otherwise $h_{\bar e}(t/n)=\infty$ for some
$t>0$ and every $n$. A convergent subsequence of levels representing
$t/n$ would give threshold zero at $e_*$ and infinity at $\bar e$,
contradicting the shared zero thresholds.

Thus $h_{\bar e}$ is finite, nonnegative and additive, so $h_{\bar e}(t)=w(\bar e)t$.
The coefficient is positive: if it were zero, a limit of levels with
reference thresholds $n\to\infty$ would have threshold infinity at
$e_*$ and zero at $\bar e$, again contradicting the shared zeros. Monotonicity
then also gives $A_{\bar e}(\lambda)=\infty$ whenever
$A_{e_*}(\lambda)=\infty$.

Finally, choose a level with reference threshold one. Its maps are
$\min\{\bar q,w(\bar e)\}$. For $\bar e_n\to \bar e$, fix $\bar q>w(\bar e)$. Joint continuity
implies $\min\{\bar q,w(\bar e_n)\}\to w(\bar e)<\bar q$, so eventually $w(\bar e_n)<\bar q$
and $w(\bar e_n)\to w(\bar e)$. Hence $w$ is continuous.
\end{proof}

\begin{proof}[Theorem~\ref{thm:type-comppair}]
Take the Young representation of the rule, calibrated at equal equity profiles.
Lemma~\ref{lem:no-mixed-cases} states that they are of finite log-exponential form, CEL form or CEA form at each profile.
Lemma~\ref{lem:finite-sync} gives the canonical representation in
the finite case on the full domain.
Lemma~\ref{lem:cea-sync} identifies the CEA case, after using
$A_{e_*}$ as the level; duality identifies CEL. These gain maps determine
the allocation at every profile by the parametric representation,
including when equities differ. This proves necessity.

Conversely, all three families have continuous Young representations,
so Proposition~\ref{prop:young} gives continuity, equal treatment and
consistency. For the finite rules,
\[
 \begin{aligned}
 F_\mu(F_\lambda(\bar q,\bar e),\bar e)&=F_{\lambda\mu}(\bar q,\bar e),\\
 \bar q-F_\lambda(\bar q,\bar e)&=\varphi_{-b(\bar e)}^{-1}((1-\lambda)\varphi_{-b(\bar e)}(\bar q)).
 \end{aligned}
\]
Thus gain maps compose with parameter $\lambda\mu$, and loss
maps with parameter $\lambda+\mu-\lambda\mu$, independently of equity.
For weighted CEA, residual caps compose by taking the smaller level,
and loss thresholds by adding levels. Weighted CEL is the dual.
Composing maps therefore reproduces the direct allocation at its
resulting total, proving DPI and UPI in every case.
\end{proof}

\section{Independence of the characterisation axioms}\label{app:independence}

Each rule below is defined for every finite population and satisfies
four of the five axioms in Theorem~\ref{thm:type-comppair}, while
violating the fifth.

\paragraph{Without equal treatment.}
Fix a strict order of account labels, inherited by every subpopulation,
and allocate positions in that order:
\[
 r_i(q,e,\widehat E)
 =\min\left\{q_i,\left(\widehat E-\sum_{j\prec i}q_j\right)_+\right\}.
\]
This rule is continuous and consistent: removing accounts with their
awards preserves the order in which the remaining positions are filled.
A smaller target truncates the same ordered allocation, giving DPI;
allocating an additional target from the unfilled positions continues
that allocation, giving UPI. For two identical types with position one
and target one, the first account receives one and the second zero.
Thus equal treatment fails.

\paragraph{Without continuity.}
Use weighted CEA with
\[
 w(x)=\begin{cases}1,&x\le1,\\2,&x>1.\end{cases}
\]
Equal types have equal weights, and a clearing level remains valid on
each reduced problem, so equal treatment and consistency hold. At fixed
equities, cap maps compose by taking the smaller level and their loss
maps compose by adding thresholds; hence DPI and UPI hold without
continuity of $w$. With $q=(1,1)$ and target one, equities $(1,1)$ give
$(1/2,1/2)$, whereas equities $(1+\varepsilon,1)$ give $(2/3,1/3)$ for
every $\varepsilon>0$. Continuity fails.

\paragraph{Without consistency.}
Use ordinary CEA for odd population sizes and proportional division
for even population sizes. Each rule separately satisfies continuity,
equal treatment, DPI and UPI, and target revisions do not change the
population. All four axioms therefore hold. For three accounts with
$q=(1,2,3)$ and target three, the allocation is $(1,1,1)$. Restricting
to the last two accounts with their assigned total two instead gives
$(4/5,6/5)$ under proportional division. Thus consistency fails.

\paragraph{Without UPI.}
Let $r$ be the type-parametric rule with schedules
\[
 f((q_0,e_0),\lambda)=\lambda^{e_0}q_0,\qquad 0\le\lambda\le1.
\]
These schedules are jointly continuous, including at $\lambda=0$ since
$e_0>0$. They give continuity, consistency and equal treatment.
Composing gain maps multiplies their levels, giving DPI. To see the
failure of UPI, take $q=(1,1)$, $e=(1,2)$ and target $3/4$. The clearing
level $1/2$ gives $(1/2,1/4)$. Allocating an additional $7/16$ from the
removed positions $(1/2,3/4)$ at level $1/2$ gives $(1/4,3/16)$.
The combined allocation is $(3/4,7/16)$ at target $19/16$, but direct
allocation from $q=(1,1)$ must have the form $(\lambda,\lambda^2)$.
The combined allocation has no such form, so UPI fails.

\paragraph{Without DPI.}
Take the dual of the preceding rule,
\[
 r^*(q,e,\widehat E)=q-r\left(q,e,\sum_iq_i-\widehat E\right).
\]
Duality preserves continuity, consistency and equal treatment and
interchanges DPI and UPI. Thus $r^*$ satisfies UPI and violates DPI.
Explicitly, on the same types, it gives $(1/2,3/4)$ at target $5/4$.
Applying it to these residuals at target $13/16$ gives $(1/4,9/16)$.
Direct allocation at $13/16$ must have the form
$(1-\lambda,1-\lambda^2)$, which $(1/4,9/16)$ does not have.

\section{Proofs for homogeneity and account manipulation}\label{app:splitting}

\begin{proof}[Proposition~\ref{prop:homogeneous}]
Rescale a problem by $t$ and divide its allocation by $t$. In the finite
case, the resulting rule has profile $\widetilde b_t(x)=t b(tx)$,
since
\[
 \varphi_{b(tx)}(tq)=t\varphi_{t b(tx)}(q).
\]
Homogeneity is therefore equivalent to $t b(tx)=b(x)$: necessity follows
by restricting to a common equity and using uniqueness in the scalar
characterisation. Taking $x=1$ gives $b(t)=b(1)/t$, and the displayed
identity also proves sufficiency. Set $\theta=b(1)/p_\tau$.

For either boundary family, rescaling replaces $w(x)$ by $w(tx)/t$.
Two positive weight profiles define the same CEA rule exactly when
they are proportional: a sufficiently small target for any pair of
accounts reveals their weight ratio. Duality gives the same statement
for CEL. Hence homogeneity is equivalent to
\[
 w(tx)=d(t)w(x)\qquad(t,x>0)
\]
for some $d(t)>0$. Taking $x=1$ shows that $A(x)=w(x)/w(1)$ is positive,
continuous and multiplicative. Thus $u\mapsto\log A(e^u)$ is continuous
and additive, giving $A(x)=x^\gamma$. Conversely, for $w(x)=Cx^\gamma$, the
level $t^{1-\gamma}\lambda$ in the rescaled problem gives exactly $t$ times
the original allocation.
\end{proof}

\begin{proof}[Proposition~\ref{prop:splitmerge}]
We use Lemma~\ref{lem:typeadd}. Fix an interior level $\lambda$.
For $q>0$, the residual is strictly increasing in $b$: $\varphi_b^{-1}(\lambda\varphi_b(q))/q$ is
the secant slope from zero of the strictly convex function
$u\mapsto\log(1-\lambda+\lambda e^u)$. Also
\[
 \frac{\partial^2}{\partial q^2}\bigl[\varphi_b^{-1}(\lambda\varphi_b(q))\bigr]
 =\frac{b\lambda(1-\lambda)e^{bq}}
        {(1-\lambda+\lambda e^{bq})^2}.
\]
Thus, as a function of $q$, it is convex for $b>0$, linear for $b=0$,
and concave for $b<0$. Since it vanishes at zero, it is respectively
superadditive, additive and subadditive.

\emph{Finite rules.}
Suppose every residual map is superadditive. Set the second position
to zero, while retaining its positive equity. Then
\[
 \varphi_{b(e_1+e_2)}^{-1}(\lambda\varphi_{b(e_1+e_2)}(q))\ge\varphi_{b(e_1)}^{-1}(\lambda\varphi_{b(e_1)}(q)).
\]
Strict monotonicity in $b$ gives $b(e_1+e_2)\ge b(e_1)$, so $b$ is
nondecreasing. If $b(e_1+e_2)<0$, all three relevant parameters are
negative. Divide superadditivity by $\lambda$ and let
$\lambda\downarrow0$, using
\[
 \frac{\varphi_b^{-1}(\lambda\varphi_b(q))}{\lambda}\longrightarrow
 \frac{e^{bq}-1}{b}.
\]
Letting both positions tend to infinity gives (S2).

Conversely, assume (S1)--(S2), and let $b_{12}=b(e_1+e_2)$.
If $b_{12}\ge0$, superadditivity at $b_{12}$ and monotonicity
in $b$ give
\begin{align*}
 \varphi_{b_{12}}^{-1}(\lambda\varphi_{b_{12}}(q_1+q_2))\ge\sum_{i=1}^2\varphi_{b_{12}}^{-1}(\lambda\varphi_{b_{12}}(q_i))\ge\sum_{i=1}^2\varphi_{b(e_i)}^{-1}(\lambda\varphi_{b(e_i)}(q_i)).
\end{align*}
If $b_{12}<0$, write $\alpha(x)=-1/b(x)$ at the three equities.
The map
\[
 \mathcal F_\lambda(q,\alpha)=-\alpha\log(1-\lambda+\lambda e^{-q/\alpha})
\]
is the perspective of a concave function. It is therefore jointly
concave and positively homogeneous, hence superadditive. It is also
increasing in $\alpha$. Condition (S2) now gives
\begin{align*}
 F_\lambda(q_1+q_2,e_1+e_2)
 &\ge \mathcal F_\lambda(q_1+q_2,\alpha(e_1)+\alpha(e_2))\\
 &\ge \mathcal F_\lambda(q_1,\alpha(e_1))+\mathcal F_\lambda(q_2,\alpha(e_2)).
\end{align*}
This proves split-proofness. Since
$q-F_\lambda(q,x)=\varphi_{-b(x)}^{-1}((1-\lambda)\varphi_{-b(x)}(q))$ and $q$ is additive,
subadditivity for $b$ is superadditivity for $-b$. The merging
conditions follow by changing the sign of $b$.

\emph{Boundary rules.}
For CEA, the map $(q,A)\mapsto\min\{q,A\}$ is concave and positively
homogeneous, hence superadditive. Superadditivity of $w$ is therefore
sufficient; taking both positions above all caps proves necessity.
For subadditivity, setting one position to zero and the other above
all caps gives
$w(e_1+e_2)\le\min\{w(e_1),w(e_2)\}$, which is equivalent to $w$
being nonincreasing. Conversely, with
$A=w(e_1+e_2)\lambda\le A_i=w(e_i)\lambda$,
\[
 \min\{q_1+q_2,A\}
 \le\min\{q_1,A\}+\min\{q_2,A\}
 \le\min\{q_1,A_1\}+\min\{q_2,A_2\}.
\]
CEL is dual to CEA, so its two conditions are interchanged.

Finally, a finite profile satisfying both conditions is constant.
A negative constant violates (S2), and a positive one violates
(M2), leaving only $b=0$. A positive weight profile cannot be both
superadditive and nonincreasing, since that would require
$2w(x)\le w(2x)\le w(x)$. Thus proportional division is the only
rule satisfying both properties.
\end{proof}

\begin{proof}[Corollary~\ref{cor:position-only}]
For a constant negative $b$, (S2) would require
$1/(-b)\ge2/(-b)$, so split-proofness requires $b\ge0$;
these nonnegative constants satisfy the splitting conditions. Duality
gives the merging statement. A positive constant weight profile is
nonincreasing but not superadditive, giving the boundary cases.
\end{proof}

\begin{proof}[Corollary~\ref{cor:homogeneous}]
For $\theta<0$, the profile $\theta p_\tau/x$ is increasing and $1/(-b(x))=x/(-\theta p_\tau)$
is additive. For $\theta>0$, it is decreasing and $1/b(x)=x/(\theta p_\tau)$ is additive.
The zero profile satisfies both manipulation conditions. Applying
Proposition~\ref{prop:splitmerge} gives the finite case. For the
boundaries, $e^\gamma$ is superadditive exactly when $\gamma\ge1$, and is
nonincreasing exactly when $\gamma\le0$.
\end{proof}

\section{Proof of Theorem~\ref{thm:fairness-rays}}\label{app:fairness}

\begin{proof}[Theorem~\ref{thm:fairness-rays}]
Fix a problem $(q,e,\widehat E)$ and abbreviate the selected allocations by $r^\kappa=r^\kappa(q,e,\widehat E)$.
At an interior target, let $\lambda=\lambda(\kappa)\in(0,1)$ clear,
and write retention fractions as
\[
 t_i=\frac{r_i^\kappa}{q_i}=G(u_i,\lambda),\qquad
 u_i=\kappa b(e_i)q_i,\qquad
 G(u,\lambda)=\frac{\log(1-\lambda+\lambda e^u)}{u},
\]
with $G(0,\lambda)=\lambda$. We show that increasing $\kappa$
transfers residual position towards accounts with higher retention
fractions, making those fractions more dispersed.

The function $G$ is strictly increasing in $u$ and satisfies
$\partial_\lambda G>0$. To compare how retention fractions change with $\kappa$, define
\[
 J_\lambda(u)=\frac{u\,\partial_uG(u,\lambda)}
                         {\partial_\lambda G(u,\lambda)}.
\]
It too is strictly increasing. Indeed, with $s=e^u$ and
$A=1-\lambda+\lambda s$,
\[
 J_\lambda(u)=\frac{\lambda s\log s-A\log A}{s-1},\qquad
 \frac{d}{ds}J_\lambda(\log s)
 =\frac{\log A-\lambda\log s}{(s-1)^2}>0.
\]
Strict concavity of the logarithm gives the inequality for $s\ne1$;
at $s=1$ the derivative has limit $\lambda(1-\lambda)/2>0$.
Also $J_\lambda(0)=0$.

For $\kappa>0$, set
\[
 v_i=q_i\partial_\lambda G(u_i,\lambda),\qquad
 J_i=J_\lambda(u_i),\qquad
 \overline J=\frac{\sum_i v_iJ_i}{\sum_i v_i}.
\]
Differentiating the schedules and the budget equation gives
\begin{equation}\label{eq:fairness-derivative}
 \dot r_i=\frac{v_i}{\kappa}(J_i-\overline J).
\end{equation}
Since $t_i$ and $J_i$ have the same order, gains go to accounts with
higher retention fractions. More explicitly, for differentiable convex
$\psi$,
\[
 \frac{d}{d\kappa}\sum_iq_i\psi(t_i)
 =\frac{\sum_{i,j}v_iv_j
       [\psi'(t_i)-\psi'(t_j)](J_i-J_j)}
        {2\kappa\sum_i v_i}\ge0.
\]
For $\psi(t)=t^2$, the inequality is strict unless all scores
$b(e_i)q_i$ agree. Integration and smooth convex approximation give
the comparison for every convex $\psi$. Take $\psi(t)=\Phi(1-t)$
to obtain the comparison. Continuity includes $\kappa=0$; extreme targets
have the same allocation under every rule.

\end{proof}

\subsection{Fairness comparisons at the homogeneous boundaries}

We verify the non-comparisons in
Remark~\ref{prop:boundary-comparison}. Fix distinct equities and $c_0>0$.
For $\mathrm{CEA}_\gamma$ with $\gamma>1$, choose $q_i=c_0e_i^\gamma$ and a sufficiently
small positive target. CEA then retains the same fraction of both
positions, whereas water-filling does not, since their initial leverages
differ. Proposition~\ref{prop:prop-fairest} makes CEA strictly fairer for
every strictly convex burden criterion. With $q_i=c_0e_i$ instead and the
target again small enough to leave both CEA residuals below their
positions, water-filling is proportional and CEA is not. This reverses
the strict comparison.

For $\mathrm{CEL}_\gamma$ with $\gamma\le0$, take $q_i=c_0e_i^\gamma$ and a positive
reduction small enough to leave both accounts with positive residuals.
CEL gives the same percentage reduction to both accounts; water-filling
does not. With $q_i=c_0e_i$ instead, water-filling is proportional while
CEL's percentage reductions are proportional to $e_i^{\gamma-1}$ and differ.
Proposition~\ref{prop:prop-fairest} again gives the two strict comparisons.

\section{Equal-risk invariance}\label{sec:risk-contributions}

Risk considerations can justify unequal percentage position reductions. When two accounts contribute the same risk per unit of equity, however, this justification provides no reason to favour one over the other. We formulate a strong and a weak version of this principle. Equal-risk invariance requires the two accounts to bear the same reduction per unit of equity in every problem. No persistent favouritism requires any unequal treatment of equal-risk types to reverse in some other problem. Within the characterised class, both requirements select the family $r^\theta$. Homogeneity follows from either requirement.

\subsection{Risk per unit of equity}

Fix the terminal price $P$ and a spectral risk measure $\rho$ satisfying the assumptions of Theorem~\ref{thm:risk-optimal-composition}. In particular, $P$ has a density positive on $(p_\tau,\infty)$, and the risk of every finite profile is finite. For an account of type $(q,e)$, define its \textbf{uncovered loss before deleveraging} and its \textbf{risk per unit of equity} by
\[
 X(q,e)=\bigl((P-p_\tau)q-e\bigr)_+,
 \qquad
 K(q,e)=\frac{\rho(X(q,e))}{e}.
\]
The contribution is measured before deleveraging, so it depends only on the account's type and the fixed risk model.

\begin{lemma}[Account risk contributions]\label{lem:account-risk}
For every finite profile,
\[
 \rho\!\left(\sum_i X(q_i,e_i)\right)
 =\sum_i e_iK(q_i,e_i).
\]
Moreover,
\[
 K(q,e)=H(q/e),
 \qquad
 H(z)=\rho\!\left(\bigl((P-p_\tau)z-1\bigr)_+\right),
\]
where $H$ is strictly increasing on $\R_+$. Consequently,
\[
 K(q,e)=K(q',e')
 \quad\Longleftrightarrow\quad
 \frac qe=\frac{q'}{e'}.
\]
\end{lemma}

\begin{proof}
Each account's loss is nondecreasing in the same terminal price. The losses are therefore comonotone, and their CVaRs add~\citep[Section~2.5]{campbell}. Integrating over the spectral weights gives the first identity. Positive homogeneity of $\rho$ gives $K(q,e)=H(q/e)$.

For strict monotonicity, let $\pi(u)$ be the quantile function of $P$. For every $0\le\beta<1$,
\[
 H_\beta(z):=\operatorname{CVaR}_\beta
       \!\left(\bigl((P-p_\tau)z-1\bigr)_+\right)
 =\frac{1}{1-\beta}\int_\beta^1
       \bigl((\pi(u)-p_\tau)z-1\bigr)_+\,du.
\]
If $0\le z_1<z_2$, the integrand for $z_2$ is no smaller everywhere and is strictly larger whenever $\pi(u)>p_\tau+1/z_2$. The price-support assumption ensures that this strict inequality holds on a set of positive measure in every upper quantile interval $(\beta,1)$. Hence $H_\beta(z_2)>H_\beta(z_1)$ for every $\beta<1$. Integrating against the probability measure $\omega$ proves strict monotonicity of $H$.
\end{proof}

Thus $e_iK(q_i,e_i)$ is an additive contribution to platform risk. Equal risk per unit of equity also has a stronger interpretation here: the two accounts have identical uncovered losses per unit of equity in every price state. If both retain the same fraction of their positions, this identity continues to hold after deleveraging.

The support assumption matters for this identification. If prices are bounded above, two different positive position-to-equity ratios may both produce zero uncovered loss. Equal risk scores would then conceal different exposures. Indeed, at the zero target their reductions per unit of equity differ. The requirements below are imposed under the fixed risk model specified above.

\subsection{Treatment of equal-risk accounts}

We compare residual positions per unit of equity. By Lemma~\ref{lem:account-risk}, accounts with equal risk per unit of equity have the same initial position-to-equity ratio. Equal residual positions per unit of equity also mean equal reductions per unit of equity.

\medskip\noindent\textbf{Equal-risk invariance.}
For every problem $(q,e,\widehat E)$ and every pair of accounts $i,j$,
\[
 K(q_i,e_i)=K(q_j,e_j)
 \quad\Longrightarrow\quad
 \frac{r_i(q,e,\widehat E)}{e_i}
 =\frac{r_j(q,e,\widehat E)}{e_j}.
\]

A weaker requirement allows unequal treatment in individual problems but requires that any preference between equal-risk types can reverse when the target or the other accounts change.

\medskip\noindent\textbf{No persistent favouritism.}
For every pair of accounts $i,j$ and types $(q_i,e_i),(q_j,e_j)$ for these accounts with
$K(q_i,e_i)=K(q_j,e_j)$, if $i$ and $j$ receive unequal residual positions per unit of equity in some problem $(q,e,\widehat E)$,
\[
 \frac{r_i(q,e,\widehat E)}{e_i}
 >
 \frac{r_j(q,e,\widehat E)}{e_j},
\]
there is a problem $(\widetilde q,\widetilde e,\widehat E')$ containing these accounts with the same types $(\widetilde q_k,\widetilde e_k)=(q_k,e_k)$ for $k=i,j$ such that
\[
 \frac{r_i(\widetilde q,\widetilde e,\widehat E')}{e_i}
 <
 \frac{r_j(\widetilde q,\widetilde e,\widehat E')}{e_j}.
\]

Equal-risk invariance implies no persistent favouritism: every pair of equal-risk types is treated equally in every problem. The converse need not hold for arbitrary rules, because the ranking of two accounts can change with the target or the surrounding profile. In the extended log-exponential family, the relevant ranking cannot reverse.

In this one-asset model, Lemma~\ref{lem:account-risk} gives
$K(q,e)=H(q/e)$ with $H$ strictly increasing. Both axioms therefore
compare precisely accounts with equal leverage. The choice of $\rho$
and $P$ affects the numerical risk scores but not which accounts are
compared, so it drops out of the rule selection below under the stated
assumptions.

\begin{proposition}[Account risk and rule selection]\label{prop:account-risk-selection}
Within the extended log-exponential family, the following are equivalent:
\begin{enumerate}
\item the rule satisfies equal-risk invariance;
\item the rule satisfies no persistent favouritism;
\item the rule belongs to the family $r^\theta$, $\theta\in[-\infty,+\infty]$.
\end{enumerate}
Equivalently, a finite rule satisfies either requirement if and only if there exists $\theta\in\R$ such that $b(x)=\theta p_\tau/x$ for every $x>0$. A boundary rule satisfies either requirement if and only if it admits the weight profile $w(x)=x$.
\end{proposition}

\begin{proof}[Proposition~\ref{prop:account-risk-selection}]
The first condition implies the second. We show that the second implies the third and that the third implies the first.

\emph{Finite rules.}
Let $k(x)=xb(x)$. By Lemma~\ref{lem:account-risk}, accounts of positive equal risk per unit of equity have positions $q_i=ze_i$ for a common $z>0$. At their common clearing level,
\[
 \frac{r_i(q,e,\widehat E)}{e_i}
 =\varphi_{k(e_i)}^{-1}(\lambda\varphi_{k(e_i)}(z)).
\]
For fixed $z>0$ and $0<\lambda<1$, $\varphi_k^{-1}(\lambda\varphi_k(z))$ is strictly increasing in $k$. To see this, $\varphi_k^{-1}(\lambda\varphi_k(z))/z$ is the secant slope from zero of the strictly convex function
\[
 u\longmapsto\log(1-\lambda+\lambda e^u),
\]
evaluated at $u=kz$, with its continuous extension at zero.

If $k(e)>k(e')$, fix any $z>0$ and take the types $(ze,e)$ and $(ze',e')$. The first account has a strictly larger residual position per unit of equity in every problem with an interior target, independently of the other accounts. At the two extreme targets the residual positions per unit of equity agree. The ranking cannot reverse, violating no persistent favouritism. Thus $k$ must be constant. Conversely, when $k(e)=\theta p_\tau$ for every $e$, the displayed formula gives equal residual positions per unit of equity for equal-risk accounts in every problem, including the extreme targets.

\emph{Boundary rules.}
Let $v(x)=w(x)/x$. For equal-risk accounts with $q_i=ze_i$,
\[
 \begin{aligned}
\frac{ \mathrm{CEA}_{w,i}(q,e,\widehat E)}{e_i}=\min\{z,\lambda v(e_i)\}\,\text{ and }\,
 \frac{ \mathrm{CEL}_{w,i}(q,e,\widehat E)}{e_i}=\max\{0,z-\lambda v(e_i)\}.
 \end{aligned}
\]
Under CEA, $v(e)>v(e')$ gives a weakly larger residual position per unit of equity to the first type in every problem containing the pair. Choose a two-account problem and $0<\lambda<z/\max\{v(e),v(e')\}$, and set the target equal to the sum of the corresponding residuals. Both caps are inactive, so the comparison is strict. Under CEL, $v(e)<v(e')$ gives the same persistent advantage to the first type; choosing $\lambda$ in the same interval keeps both residuals positive and makes the comparison strict. Hence no persistent favouritism requires $v$ to be constant. A common positive factor in $w$ leaves the rule unchanged, so the rule admits $w(x)=x$. This profile gives equal residual positions per unit of equity whenever risks per unit of equity coincide.

Finally, $K(q,e)=0$ implies $q=0$ by Lemma~\ref{lem:account-risk}. Such accounts have zero residual positions under every feasible rule and impose no further restriction.
\end{proof}

For equal-risk accounts with positive positions,
$r_i(q,e,\widehat E)/e_i=z\,r_i(q,e,\widehat E)/q_i$ with the same $z=q_i/e_i$.
Equal residual positions per unit of equity therefore also mean equal percentage reductions for the accounts being compared. The requirements leave the parameter $\theta$ unrestricted. Among the remaining homogeneous boundary rules, the split-proof members favour the larger-equity account whenever two accounts with equal risk per unit of equity receive different treatment. This preference never reverses as the target or the other accounts change.

\section{Proofs for Section~\ref{sec:risk}}\label{app:risk-details}

\subsection{Proofs of the main risk comparisons}

\begin{proof}[Lemma~\ref{lem:risk-order}]
Equal targets give equal equity-weighted mean leverage. On the
finite union of the two leverage supports, every convex function agrees
with an affine function plus a nonnegative linear combination of hinges
$(z-k)_+$. Hence
\[
 r\preceq_e s
 \quad\Longleftrightarrow\quad
 \sum_i(r_i-ke_i)_+\le\sum_i(s_i-ke_i)_+
 \quad(k\ge0).
\]
Here and below, $x_+=\max\{0,x\}$. For $d=p-p_\tau>0$,
\[
 L(r,p)=d\sum_i(r_i-e_i/d)_+.
\]
As $p$ ranges over $(p_\tau,\infty)$, $1/d$ ranges over $(0,\infty)$.
The hinge inequality at zero is equality, and both losses vanish for
$p\le p_\tau$. This proves the equivalence. Applying monotonicity of
$\rho$ to the resulting almost-sure loss inequality proves the last claim.
\end{proof}

\begin{proof}[Proposition~\ref{prop:risk-monotonicity}]
Fix a problem $(q,e,\widehat E)$ and abbreviate the selected allocations by $r^\theta=r^\theta(q,e,\widehat E)$.
The scalar result in \citet[Proposition~2.1]{schlegel} establishes the scalar Lorenz
ranking, including its endpoints. To obtain equity weights, first write
$e_i=hk_i$ with $h>0$ and positive integers $k_i$. Replace account $i$
by $k_i$ scalar claimants with claim $q_i/e_i$, and use estate
$\widehat E/h$. Each receives $r_i^\theta/e_i$ under the scalar rule with parameter
$\theta p_\tau$. Multiplying the scalar
comparison by $h$ gives
\[
 \sum_i e_i\Phi(r_i^{\theta_1}/e_i)
 \le\sum_i e_i\Phi(r_i^{\theta_2}/e_i).
\]
Approximate arbitrary positive equities by commensurable ones and use
continuity of the clearing allocations. Lemma~\ref{lem:risk-order}
then gives the risk comparison.
\end{proof}

\begin{proof}[Proposition~\ref{prop:waterfill-risk}]
Write $r_i^{\mathrm{WF}}=\min\{q_i,e_i\lambda\}$. We compare hinges,
as in Lemma~\ref{lem:risk-order}. If $k\ge\lambda$, then
$\sum_i(r_i^{\mathrm{WF}}-ke_i)_+=0$. If $0\le k<\lambda$, then for
any feasible $s$,
\[
 \sum_i(r_i^{\mathrm{WF}}-ke_i)_+
 =\widehat E-\sum_i\min\{q_i,ke_i\}
 \le\widehat E-\sum_i\min\{s_i,ke_i\}
 =\sum_i(s_i-ke_i)_+.
\]
Thus $r^{\mathrm{WF}}\preceq_e s$.
\end{proof}

\subsection{Proof of universal risk minimisation}

\begin{proof}[Theorem~\ref{thm:risk-optimal-composition}]
The key is strict convexity at sufficiently small leverages. Let
$\pi(u)$ be the quantile function of $P$. All account losses are nondecreasing in the same price $P$, so their upper-tail averages use the same price quantiles. Their CVaRs therefore add, and the objective is $\sum_i e_iH(r_i/e_i)$, with the function $H$ from Lemma~\ref{lem:account-risk}:
\[
 H(z)=\int_{[0,1)}H_\beta(z)\,\omega(d\beta),\qquad
 H_\beta(z)=\frac{1}{1-\beta}\int_\beta^1
                \bigl((\pi(u)-p_\tau)z-1\bigr)_+\,du.
\]
Each $H_\beta$ is convex. Choose $0<\beta_0<1$ with
$\omega([0,\beta_0])>0$, and $\eta>0$ so small that
$p_\tau+1/\eta>\pi(\beta_0)$. For $0\le z_1<z_2\le\eta$, prices in
\[
 (p_\tau+1/z_2,\ p_\tau+1/z_1)
\]
have positive probability and lie above the $\beta_0$-quantile; the
right endpoint is $\infty$ when $z_1=0$. At these prices the hinge is
strictly below the chord between $z_1$ and $z_2$. Thus every
$H_\beta$ with $\beta\le\beta_0$ is strictly convex on $[0,\eta]$,
and so is $H$. Water-filling is consequently the unique optimum
whenever all initial leverages are at most $p_\tau\eta$.

By Theorem~\ref{thm:type-comppair}, only three families need be considered.
With equities $(1,1)$, positions $(\varepsilon,3\varepsilon)$ and target
$3\varepsilon$, where $3\varepsilon<\eta$, the unique optimum is
$(\varepsilon,2\varepsilon)$. Finite log-exponential rules are strictly
interior and cannot fully retain the first position. Weighted CEL also
reduces every positive position at an interior target. Hence a universal
minimiser must be weighted CEA.

Fix any two equities and set $q_i=e_i\eta/2$. At sufficiently small
positive targets, neither the CEA allocation nor water-filling meets a
position cap. Uniqueness therefore requires
$w(e_1)\lambda/e_1=w(e_2)\lambda/e_2$. Since the equities were arbitrary,
$w(x)=Cx$ for a common $C>0$. This is water-filling, whose optimality
follows from Proposition~\ref{prop:waterfill-risk}.
\end{proof}

\subsection{Risk comparisons at the homogeneous boundaries}\label{app:boundary-risk}

We verify the non-comparisons in Remark~\ref{rem:homogeneous-boundary-risk}. Fix $\gamma\ne1$ and a finite parameter $\theta$. First choose two distinct equities and equal positive initial leverages, $q_i=\ell e_i/p_\tau$. Every finite $r^\theta$ gives equal surviving leverages. For $\mathrm{CEA}_\gamma$, choose a sufficiently small positive target so that neither position cap binds. The surviving leverages are then $p_\tau\lambda e_i^{\gamma-1}$ and differ. For $\mathrm{CEL}_\gamma$, choose a sufficiently small positive reduction so that both residuals remain positive; their leverages are $\ell-p_\tau\lambda e_i^{\gamma-1}$ and again differ. Jensen's inequality shows that $r^\theta$ is strictly lower in the risk order in both cases.

For the reverse CEA comparison, take $e=(1,1)$, $q=(1,3)$ and $\widehat E=2$. Every $\mathrm{CEA}_\gamma$ gives $(1,1)$, while a finite $r^\theta$ strictly reduces the first position and therefore leaves the second residual above one. For the reverse CEL comparison, choose two distinct equities, any $t>0$, and positions $q_i=te_i+e_i^\gamma$, with target $\widehat E=t(e_1+e_2)$. Rule $\mathrm{CEL}_\gamma$ clears at $\lambda=1$ and gives $r_i=te_i$. Under finite $r^\theta$, surviving leverage is $\varphi_\theta^{-1}(\lambda\varphi_\theta(p_\tau(t+e_i^{\gamma-1})))$, which differs across the accounts because $\varphi_\theta^{-1}(\lambda\varphi_\theta(\cdot))$ is strictly increasing at every interior level. In each case the boundary rule equalises surviving leverage and is strictly lower in the risk order. Strictness means that the reverse order fails, or equivalently that uncovered loss is strictly smaller at some terminal price.

\section{Proof of Proposition~\ref{prop:universal-risk}}\label{app:risk-comparisons}

\begin{proof}[Proposition~\ref{prop:universal-risk}]
Identical profiles give identical allocations, and
Proposition~\ref{prop:risk-monotonicity} gives sufficiency for the
homogeneous profiles. For necessity, let $k_j(x)=x b_j(x)$. We choose
a two-account problem on which rule 2 has constant surviving leverage;
rule 1 must then give the same allocation.

Fix equities $e,f>0$. For sufficiently small $z>0$, set
\[
 \ell_h(z)=\frac{\log(2 \exp(k_2(h)z)-1)}{k_2(h)},\qquad
 q_h=h\ell_h(z),\qquad h\in\{e,f\},
\]
where $\ell_h(z)=2z$ if $k_2(h)=0$. At target $(e+f)z$, rule 2
clears at level $1/2$ and gives both accounts leverage $p_\tau z$.
The assumed convex-order comparison, applied to $(x-z)^2$, forces
rule 1 to give the same leverages. Its required levels
\[
 \lambda_1(h,z)=
 \frac{\exp(k_1(h)z)-1}{\exp(k_1(h)\ell_h(z))-1}
\]
must therefore agree for $h=e,f$; at $k_1(h)=0$, use $z/\ell_h(z)$.

Write $D(h)=k_1(h)-k_2(h)$. Expanding
$\ell_h(z)=2z-k_2(h)z^2+k_2(h)^2z^3+O(z^4)$ gives
\[
 \lambda_1(h,z)=\frac12-\frac{D(h)}4z
                  +\frac{k_2(h)D(h)}8z^2+O(z^3).
\]
The formulas extend to zero parameters by continuity. Equality of the
two levels for every small $z$ yields
\[
 D(e)=D(f),\qquad k_2(e)D(e)=k_2(f)D(f).
\]
Since the equities were arbitrary, $D$ is constant. If $D=0$, the
profiles agree. Otherwise both $k_2$ and $k_1$ are constant, so
$b_j(x)=\theta_jp_\tau/x$ with $\theta_j=k_j/p_\tau$.

To determine the order, take two equal-equity accounts with distinct
positive positions and an interior target. The scalar ranking in
\citet[Proposition~2.1]{schlegel} is strict on such a pair.
Thus $\theta_1>\theta_2$ would make rule 1 strictly more dispersed, contrary
to the hypothesis. Hence $\theta_1\le \theta_2$.
\end{proof}

\section{The heterogeneous fairness example}\label{app:heterogeneous-selection}

\begin{proof}[Example~\ref{ex:heterogeneous-fairness}]
Throughout, $p_\tau=1$ and $P=p_\tau+D$ with $D$ exponentially
distributed with mean one, and $\beta=1-e^{-2}$. Since $L(r,p)$ is nondecreasing in $p$,
\[
 \operatorname{CVaR}_\beta(L(r,P))=\mathbb E[L(r,P)\mid D\ge2];
\]
this identity uses the quantile-integral definition of CVaR and remains
valid when the loss has an atom at zero. Burden variance is
\[
 V(r;q)=\sum_i\frac{q_i}{Q}(h_i-\overline h)^2,
 \qquad \overline h=1-\frac{\widehat E}{Q},
\]
with $Q=\sum_iq_i>0$ and zero positions omitted. Minimising $V$ is
minimising $B_\Phi$ for $\Phi(h)=h^2$, since the mean reduction
is fixed.

On the price tail $D\ge2$,
$L(r,P)\ge D\widehat E-\sum_i e_i$, with equality throughout the
tail exactly when $r_i\ge e_i/2$ for all $i$. Such allocations exist,
so the minimum CVaR is $3(15/2)-6=16.5$, and these floors describe its
optimal set.

We first minimise burden dispersion over this set. Write
$\psi(t)=\Phi(1-t)$. Conditional on $r_1$, Jensen's inequality
for the second and third accounts gives
\[
 B_\Phi(r;q)
 \ge\frac34\psi\left(\frac{4r_1}{3}\right)
       +13\psi\left(\frac{15/2-r_1}{13}\right).
\]
For $r_1\ge1/2$, the first argument is at least $2/3$ and the
second at most $7/13$. Increasing $r_1$ transfers mass from the lower
to the higher retention fraction, so the right-hand side is
nondecreasing, and strictly increasing if $\psi$ is strictly convex.
Its minimum is attained at $r_1=1/2$, with equal remaining fractions
$7/13$. The resulting allocation is $r^\dagger$ and satisfies all
floors and caps. Strict Jensen gives uniqueness for every strictly
convex burden measure.

Under $b(x)=-\kappa/x^2$, accounts 2 and 3 have identical effective
claims $q_i/e_i^2=1$, so their retention fractions agree. Account 1
has effective claim $3/4$. By~\eqref{eq:fairness-derivative}, its
residual is strictly increasing in $\kappa>0$, since its score is
strictly larger than the other two. This residual runs from $9/22$
at proportional division to $15/28$ at $\mathrm{CEA}_2$. There is
therefore a unique finite $\kappa_*$ with $r_1=1/2$, and the budget
then forces $r=r^\dagger$. Numerically, $\kappa_*\approx5.30749$;
it solves
\[
 \frac{1-e^{-\kappa/2}}{1-e^{-3\kappa/4}}
 =\frac{1-e^{-7\kappa/13}}{1-e^{-\kappa}}.
\]

No homogeneous rule selects $r^\dagger$.
A finite nonzero $\theta$ gives different retention fractions to accounts
2 and 3 because their leverages differ; $\theta=0$ violates the first
risk floor. If $\mathrm{CEA}_\gamma$ selected $r^\dagger$, all positions
would be uncapped, so $r_3/r_2=(3/2)^\gamma=9/4$ would force $\gamma=2$.
But then $r_2/r_1=4$, whereas $r_2^\dagger/r_1^\dagger=56/13$.
Finally, if $\mathrm{CEL}_\gamma$ selected $r^\dagger$, all residuals
would be positive, so the buybacks $(1/4,24/13,54/13)$ would be
proportional to $e_i^\gamma$: the ratio $9/4=(3/2)^\gamma$ of the third to
the second would force $\gamma=2$. But then the ratio of the second to
the first would be $4$, whereas it is $96/13$. These cases exhaust
Proposition~\ref{prop:homogeneous}. Strict optimality of $r^\dagger$
therefore proves the fairness comparison against the entire
homogeneous class.
\end{proof}

The advantage continues to hold in a neighborhood of the example.

\begin{proposition}[A fixed rule on a neighbourhood]
\label{prop:neighbourhood-fairness}
There are a constant $\bar\kappa>0$ and an open neighbourhood
$\mathcal U$ of the three-account problem in
Example~\ref{ex:heterogeneous-fairness} such that the fixed rule
\[
 b(x)=-\frac{\bar\kappa}{x^2}
\]
minimises CVaR at every $\mathcal P\in\mathcal U$ and has strictly smaller
burden variance there than every homogeneous rule
that minimises CVaR at $\mathcal P$.
\end{proposition}

\begin{proof}[Proposition~\ref{prop:neighbourhood-fairness}]
Write $\mathcal P_0=(q^0,e^0,\widehat E_0)$ for the example's problem. We first put all relevant
homogeneous rules in a compact parameter set. By
Proposition~\ref{prop:homogeneous}, the homogeneous class consists of
the family $r^\theta$, $\theta\in[-\infty,+\infty]$, whose parameter set
is compact, and of
$\mathrm{CEA}_\gamma$ and $\mathrm{CEL}_\gamma$ for $\gamma\in\R$.
The latter two ranges can be restricted uniformly near $\mathcal P_0$.

Near $\mathcal P_0$, the inequalities $q_i>e_i/2$ and
$\sum_i e_i/2<\widehat E<\sum_iq_i$ persist, so allocations satisfying
all risk floors exist. CVaR-optimality is therefore equivalent to
$r_i\ge e_i/2$ for every account. The inequalities $e_1<e_2<e_3$,
$q_1+q_2<\widehat E$ and $Q-\widehat E<q_3$ persist as well.
For CEA, the first floor implies $\lambda e_1^\gamma\ge e_1/2$, hence
\[
 r_3\ge\min\left\{q_3,\frac{e_1}{2}
                         \left(\frac{e_3}{e_1}\right)^\gamma\right\}.
\]
At $\mathcal P_0$ this exceeds the target for every $\gamma\ge3$, and the strict
inequality persists nearby. For $\gamma\le-1$, the level
$\lambda=(\widehat E-q_1-q_2)e_3^{-\gamma}$ gives the third account
$\widehat E-q_1-q_2<q_3$ and caps the first two accounts whenever
\[
 (\widehat E-q_1-q_2)\left(\frac{e_3}{e_i}\right)^{-\gamma}\ge q_i,
 \qquad i=1,2.
\]
The left-hand sides are nondecreasing in $-\gamma$. At $\mathcal P_0$
and $\gamma=-1$, the two inequalities are $33/4>3/4$ and $33/8>4$.
Hence, near $\mathcal P_0$, every $\mathrm{CEA}_\gamma$ with
$\gamma\le-1$ gives $(q_1,q_2,\widehat E-q_1-q_2)$, as
$\mathrm{CEA}_{-1}$ does. For CEL with $\gamma\le0$ and ordered
equities $e_1<e_2<e_3$, the first floor implies
$\lambda e_1^\gamma\le q_1-e_1/2$, so every buyback is at most
$q_1-e_1/2$. Thus
\[
 Q-\widehat E\le3(q_1-e_1/2).
\]
At $\mathcal P_0$, this would require $25/4\le3/4$, and it remains impossible
nearby. For $\gamma\ge0$, the third buyback is at most
$Q-\widehat E<q_3$, so it equals $\lambda e_3^\gamma$, and the first
two buybacks are at most $(Q-\widehat E)(e_2/e_3)^\gamma$. As
$\gamma\to+\infty$, the CEL allocations therefore converge to
$(q_1,q_2,q_3-(Q-\widehat E))$, uniformly near $\mathcal P_0$; we
adjoin this limit as $\gamma=+\infty$. Therefore all CVaR-optimal
homogeneous rules near $\mathcal P_0$ are represented in the compact
union of $\theta\in[-\infty,+\infty]$, CEA exponents
$\gamma\in[-1,3]$ and CEL exponents $\gamma\in[0,+\infty]$.

Allocations are jointly continuous in each compact family and the
problem, including $\theta=\pm\infty$ by
Remark~\ref{rem:type-endpoint-limits} and $\gamma=+\infty$ by the
uniform convergence above. CVaR is continuous here too:
loss is Lipschitz in $(r,e)$ with an integrable random coefficient,
and the confidence level is fixed below one. Hence the subset
$\mathcal A$ of this compact union whose members minimise CVaR at $\mathcal P_0$ is
nonempty and compact. By the preceding proof, $r^\dagger$ uniquely
minimises variance over the risk-optimal set, and no homogeneous rule
selects it; the adjoined limit gives $(3/4,4,11/4)\ne r^\dagger$ at
$\mathcal P_0$. Consequently
\[
 \Delta=\min_{h\in \mathcal A}V(r^h(\mathcal P_0);q^0)-V(r^\dagger;q^0)>0.
\]

For the heterogeneous profile $-\kappa/x^2$, the first residual
increases strictly at $\kappa_*$. Choose
$\bar\kappa>\kappa_*$ close enough that its variance at $\mathcal P_0$ is
less than $V(r^\dagger;q^0)+\Delta/2$. All three risk floors are then
strict: the first by this increase, the other two by continuity.
They persist on a neighbourhood, where the fixed rule attains the
minimum CVaR $3\widehat E-\sum_i e_i$.

Its variance advantage also persists after shrinking the neighbourhood.
Otherwise, there would be $\mathcal P_n\to \mathcal P_0$ and CVaR-optimal homogeneous
rules $h_n$ with no greater variance. The compact restriction above
gives a convergent subsequence. Continuity makes its limit a member
of $\mathcal A$, while the limiting variance comparison contradicts the gap
$\Delta/2$. This proves both assertions on one open neighbourhood.
\end{proof}

The fixed rule retains all the axioms on the full domain; its risk and
variance guarantees hold on $\mathcal U$. For any probability
distribution supported on $\mathcal U$, it therefore has lower expected
burden variance than every homogeneous rule that
minimises CVaR throughout $\mathcal U$. This neighbourhood result concerns
variance, rather than simultaneous comparison of every convex burden
measure.

\section{Proof of Proposition~\ref{thm:entry-irrelevance}}\label{app:reopening}

\begin{proof}[Proposition~\ref{thm:entry-irrelevance}]
We first show that equal positions and equities receive equal residuals
even when entry prices differ. Fix a terminal price $p'$, two accounts
with position $q>0$ and equity $e$, and admissible entry prices $s,t$.
Starting from entries $(s,s)$, reopen the first account at price $t$.
Reopening-proofness bounds its terminal residual under $(t,s)$ by half
the target. Starting instead from $(t,s)$ and reopening the first
account at $s$ gives the reverse inequality. Both comparisons are
admissible: equity at the earlier price $v\in\{s,t\}$ is
$e+q(p'-v)>0$, exactly the positive cash margin associated with entry
$v$. Thus the two residuals agree. For $q=0$ this is immediate.

Now fix any problem and account $i$. Add an account $k$ with the same
position and equity as $i$ and another admissible entry price. DPI implies
that allocations are nondecreasing in the target; budget balance then
gives
\[
 0\le r_j(E')-r_j(E)\le E'-E\qquad(E\le E').
\]
Consequently, as the enlarged target increases from zero to total
positions, the original accounts' aggregate residual varies continuously
from zero to $\sum_{j\in N}q_j$. Choose an enlarged target at which this
aggregate is $\widehat E$. Consistency recovers the original allocation.
On the pair $\{i,k\}$, consistency and the preceding argument give
$r_k=r_i$. Removing $i$ therefore leaves the same target and allocation,
with only its entry price and label changed.

To see that the new label is immaterial, take two disjoint labelled copies of any problem at twice its target. Equal treatment assigns the same residual to corresponding accounts, so each copy receives the original target in total. Consistency recovers the allocation on each copy, proving invariance under relabelling. Repeating the replacement above therefore proves entry-price independence.

Conversely, reopening changes neither positions nor later equities, so
an entry-price-independent rule satisfies the axiom with equality.
\end{proof}

\end{document}